\documentstyle{article}
\def\newpic#1{%
   \def\emline##1##2##3##4##5##6{%
      \put(##1,##2){\special{em:point #1##3}}%
      \put(##4,##5){\special{em:point #1##6}}%
      \special{em:line #1##3,#1##6}}}
\newpic{}
\begin{document}

\begin{center}
{\large \bf
LATTICE BOLTZMANN APPROACH TO VISCOUS FLOWS BETWEEN PARALLEL PLATES}
\vspace{3em}

{B\'ELA SZIL\'AGYI \\
Department of Theoretical and Computational Physics\\
University of Timi\c soara \\
Bd. Vasile P\^arvan 4, RO -- 1900 Timi\c soara, Romania}
\vspace{2em}

{ROMEO SUSAN -- RESIGA \\
Department of Hydraulic Machinery,
Technical University of Timi\c soara \\
Bd. Mihai Viteazul 1, RO -- 1900 Timi\c soara, Romania}
\vspace{4em}

{VICTOR SOFONEA \\
Research Center for Hydrodynamics,
Cavitation and Magnetic Fluids \\
Technical University of Timi\c soara\\
Bd. Mihai Viteazul 1, RO -- 1900 Timi\c soara, Romania}

\end{center}

\begin{abstract}
Four different kinds of laminar flows between two parallel plates
are investigated using the Lattice Boltzmann Method (LBM).
The LBM accuracy is estimated in two cases using numerical fits of the
parabolic velocity profiles and the kinetic energy decay curves, respectively.
The error relative to the analytical kinematic viscosity values was found
to be less than one percent in both cases.
The LBM results for the unsteady development of the flow when one plate
is brought suddenly at a constant velocity,
are found in excellent agreement with the analytical solution.
Because the classical Schlichting's approximate solution for the
entrance--region flow is not valid for small Reynolds numbers, a
Finite Element Method solution was used in order to check the accuracy of
the LBM results in this case.
\vspace{2em}

keywords: laminar flow, parallel plates,
velocity profile, Lattice Boltzmann.
\end{abstract}

\section{Introduction}

Since Frisch, Hasslacher and Pomeau \cite{b1,b2}, have shown that particles
moving on a hexagonal lattice with very simple collisions rules on its nodes
lead to the Navier-Stokes equation, the use of lattice gas
models (LGM) to study hydrodynamics has received considerable interest.

The Lattice Boltzmann Method (LBM) \cite{b3,b4,b5},
a derivative of the lattice gas automaton method, uses real
numbers instead of bits to represent particle distributions, being far less
noisy than LGM.
The LBM is a finite-difference technique for solving the kinetic equation in
discrete space and discrete time; the Navier-Stokes equation is recovered in
the long-wavelength and low-frequency limit.

This paper presents a LBM study of four different laminar flows between two parallel
plates. Both steady and unsteady flows are simulated; the numerical
results are compared with analytical solutions or Finite Element Method (FEM)
results.

\section{Lattice Boltzmann Method and Single Time Relaxation Approximation}

In the LBM, the evolution equation for the particle distribution functions
$n_i({\vec x},t) $  corresponding to the velocities $\vec c_i$ ,
$i=0,1,\ldots 6$\, ($\vec c_0\,=\,\vec 0$,
$\vert\,\vec c_j\,\vert\,=\,1$, $j=1,2,\ldots 6$)
defined on a two dimensional hexagonal lattice \cite{b5}
can be written as follows:
\begin{equation}
n_i({\vec x+\vec c_i},t+1) = 
n_i({\vec x},t) + \Omega(n_i({\vec x},t))  \label{eq1}
\end{equation}
where $\Omega_i = \Omega(n_i({\vec x},t)) $ is the local collision operator,
depending only on the local particle distribution functions $ n_i $.

Chen et al. \cite{b5} used the single time relaxation approximation
\begin{equation}
\Omega_i = - \frac{\,n_i - n_i^{eq}\,}{\tau} \label{eq2}
\end{equation}

where the equilibrium distribution function $n_i^{eq} $  depends upon the
local fluid variables and the lattice relaxation time $ \tau $ controls
the evolution to the equilibrium state.

The total mass $n$ and local momentum $n\vec u $ are conserved after
collisions:
\begin{eqnarray}
n & = & \sum_i n_i  =  \sum_i n_i^{eq} \label{eq3} \\
n\vec u & = & \sum_i \vec c_i n_i = \sum_i \vec c_i n_i^{eq} \label{eq4} 
\end{eqnarray}
The volumic density $\rho $ for a hexagonal lattice is expressed in terms
of $n$ \cite{b6}:
\begin{equation}
\rho = \frac{n}{\sqrt{3}/2} \label{eq5}
\end{equation}
To derive the mass and momentum conservation equations, a Taylor expansion
in time and space of eq.(\ref{eq1}) is performed in the long-wavelength and
low-frequency limit. Using the Chapman-Enskog procedure, the final results are:
\begin{eqnarray}
   \frac{\partial n}{\partial t} +
   \frac{\partial  }{\partial x_{\alpha}}(n u_{\alpha}) = 0 \label{eq6} \\
   \frac{\,\partial (nu_{\alpha})\,}{\partial t} +
   \frac{\partial  }{\partial x_{\beta}}
   \left[ \sum_i c_{i\,\alpha} c_{i\,\beta} n_i^{eq} - \frac{2\tau -1}{2}
   \frac{\partial  }{\partial x_{\gamma}}
   \sum_i c_{i\,\alpha} c_{i\,\beta} c_{i\,\gamma} n_i^{eq}
   \right] = 0 \label{eq7} 
\end{eqnarray}
For the hexagonal lattice we have:
\begin{eqnarray}
\vec c_i \; \; \; & = & (c_{i\, 1}, c_{i\, 2}) = \left( \cos\frac{\pi(i-1)}{3}, \sin\frac{\pi(i-1)}{3} \right)
\quad i = 1,2\ldots 6 \label{eq8} \\
 n_i^{eq} & = & n \left[ \frac{1-d_0}{6} + \frac{1}{3}(\vec c_i\cdot\vec u) +
\frac{2}{3}(\vec c_i\cdot\vec u)^2 - \frac{u^2}{6} \right]
 \quad i = 1,2\ldots 6 \label{eq9} \\
 n_0^{eq} & = & n \left[ d_0 - u^2 \right] \label{eq10}
\end{eqnarray}
where $d_0\,\in\,(0,1)$ is a constant. This leads to:
\begin{eqnarray}
 \sum_i c_{i\,\alpha} c_{i\,\beta} n_i^{eq} =
\frac{n(1-d_0)}{2} \label{eq11} \\
 \frac{\partial  }{\partial x_{\gamma}}
   \sum_i c_{i\,\alpha} c_{i\,\beta} c_{i\,\gamma} n_i^{eq} =
\frac{1}{2} S_{\alpha\beta} + \frac{1}{4} n 
(\nabla\cdot\vec u) \delta_{\alpha\beta} \label{eq12} 
\end{eqnarray}
where $S_{\alpha\beta}= \frac{1}{2}\left(
\frac{\partial u_{\beta}}{\partial x_{\alpha}} +
\frac{\partial u_{\alpha}}{\partial x_{\beta}} \right) $
is the time-rate-of-strain tensor.

Dividing (\ref{eq7}) with $\sqrt{3}/2 $ and identifying the pressure as
$p = \rho(1-d_0)/2 $, the Navier-Stokes equation in the incompressible limit
$(\nabla\cdot\vec u=0) $ is obtained as:
\begin{equation}
 \frac{\partial (\rho\vec u)}{\partial t} +
\nabla\cdot \left( p\vec I + \rho\vec u\vec u - \frac{\,2\tau-1\,}{4}\rho\vec S
\right) = 0 \label{eq13} 
\end{equation}
from which we can identify the shear $\eta $ and kinematic $\nu $
viscosities:
\begin{equation}
\eta = \rho\,\frac{\,2\tau-1\,}{8}\quad\hbox{and}\quad
  \nu = \frac{\,2\tau-1\,}{8} \label{eq14} 
\end{equation}
At the beginning of each run, the fluid was always considered to be at rest
$(\vec u = 0) $. By choosing $\rho = 2.1/(\sqrt{3}/2) $ and $d_0 = 1/7 $,
we have an uniform particle distribution $n_i = 0.3, i = 0,1,2\ldots 6 $ at
each node.
For each time step, the following successive operations are performed at every
lattice node:

\begin{itemize}
\item the local velocity is computed from (\ref{eq4});
\item the equilibrium distribution functions $n_i^{eq}(\vec x,t) $ are computed
from (\ref{eq9}) and (\ref{eq10});
\item using (\ref{eq1}) and (\ref{eq2}), the propagation step is performed in order to obtain the
$\,$ $ n_i(\vec x+\vec c_i,t+1) $ distributions;
\item if a body force exist, the new $n_i $ distributions are modified 
according to the induced particle momentum change during one time step.
\end{itemize}

\section{Flow between parallel rest plates}

\begin{figure}
\begin{center}
\caption{\label{fig1} The flow domain and the hexagonal lattice}

\def\myline#1#2#3#4{\put(#1,#2){\special{em:moveto}}%
                    \put(#3,#4){\special{em:lineto}}}

\unitlength=1mm
\special{em:linewidth 0.4pt}
\linethickness{0.4pt}
\begin{picture}(100.00,70.00)
\put(0,43.3){\rule{70.00\unitlength}{1.00\unitlength}}
\put(0,-1){\rule{70.00\unitlength}{1.00\unitlength}}
\put(-10,0){\vector(1,0){115}}
\put(0,-10){\vector(0,1){65}}
\put(-5,-5){\makebox(0,0)[cc]{0}}
\put(95.00,-5){\makebox(0,0)[cc]{$x$}}
\put(-5,50){\makebox(0,0)[cc]{$y$}}
\multiput(0,4.33013)(10,0){7}{\circle*{1}}
\put(70,4.33013){\circle{1}}
\multiput(5,12.99038)(10,0){7}{\circle*{1}}
\put(75,12.99038){\circle{1}}
\multiput(0,21.65064)(10,0){7}{\circle*{1}}
\put(70,21.65064){\circle{1}}
\multiput(5,30.31089)(10,0){7}{\circle*{1}}
\put(75,30.31089){\circle{1}}
\multiput(0,38.97114)(10,0){7}{\circle*{1}}
\put(70,38.97114){\circle{1}}
\myline{0}{21.65013}{10}{38.97114}
\myline{0}{4.33013}{20}{38.97114}
\myline{10}{4.33013}{30}{38.97114}
\myline{20}{4.33013}{40}{38.97114}
\myline{30}{4.33013}{50}{38.97114}
\myline{40}{4.33013}{60}{38.97114}
\myline{50}{4.33013}{65}{30.31089}
\myline{60}{4.33013}{65}{12.99038}
\multiput(70,4.33013)(0.5,0.866023){11}{\circle*{0.03}}
\multiput(65,12.99038)(0.5,0.866023){21}{\circle*{0.03}}
\multiput(65,30.31089)(0.5,0.866023){11}{\circle*{0.03}}
\myline{0}{21.65064}{10}{4.33013}
\myline{0}{38.97114}{20}{4.33013}
\myline{10}{38.97114}{30}{4.33013}
\myline{20}{38.97114}{40}{4.33013}
\myline{30}{38.97114}{50}{4.33013}
\myline{40}{38.97114}{60}{4.33013}
\myline{50}{38.97114}{65}{12.99038}
\myline{60}{38.97114}{65}{30.31089}
\multiput(65,12.99038)(0.5,-0.866023){11}{\circle*{0.03}}
\multiput(65,30.31089)(0.5,-0.866023){21}{\circle*{0.03}}
\multiput(70,38.97114)(0.5,-0.866023){11}{\circle*{0.03}}
\myline{0}{4.33013}{60}{4.33013}
\myline{5}{12.99038}{65}{12.99038}
\myline{0}{21.65064}{60}{21.65064}
\myline{5}{30.31089}{65}{30.31089}
\myline{0}{38.97114}{60}{38.97114}
\multiput(60,4.33013)(1,0){11}{\circle*{0.03}}
\multiput(65,12.99038)(1,0){11}{\circle*{0.03}}
\multiput(60,21.65064)(1,0){11}{\circle*{0.03}}
\multiput(65,30.31089)(1,0){11}{\circle*{0.03}}
\multiput(60,38.97114)(1,0){11}{\circle*{0.03}}
%
%
\myline{-7}{43.3}{-1}{43.3}
\put(-5,23.8){\vector(0,1){19.5}}
\put(-5,19.5){\vector(0,-1){19.5}}
\put(-5,21.65){\makebox(0,0)[cc]{$H$}}
\myline{70}{-2}{70}{-9}
\put(33,-7){\vector(-1,0){33}}
\put(37,-7){\vector(1,0){33}}
\put(35,-7){\makebox(0,0)[cc]{$L$}}
\myline{72}{4.33}{84}{4.33}
\put(82.5,-4){\vector(0,1){4}}
\put(82.5,8.33){\vector(0,-1){4}}
\put(87,4.33){\makebox(0,0)[cc]{$\frac{\sqrt{3}}{4}$}}
\myline{72}{38.97114}{84}{38.97114}
\myline{72}{43.30127}{84}{43.30127}
\put(82.5,34.97114){\vector(0,1){4}}
\put(82.5,47.30127){\vector(0,-1){4}}
\put(87,43.3){\makebox(0,0)[cc]{$\frac{\sqrt{3}}{4}$}}
\myline{77}{30.31089}{84}{30.31089}
\myline{72}{21.65064}{84}{21.65064}
\put(82.5,25.31089){\vector(0,1){5}}
\put(82.5,26.65064){\vector(0,-1){5}}
\put(87,25.53){\makebox(0,0)[cc]{$\frac{\sqrt{3}}{2}$}}
\myline{50}{4.33012}{50}{-5}
\myline{60}{4.33012}{60}{-5}
\put(56,-4){\vector(1,0){4}}
\put(54,-4){\vector(-1,0){4}}
\put(55,-4){\makebox(0,0)[cc]{{\small 1}}}
\end{picture}
\vspace{7ex}
\end{center}
\end{figure}
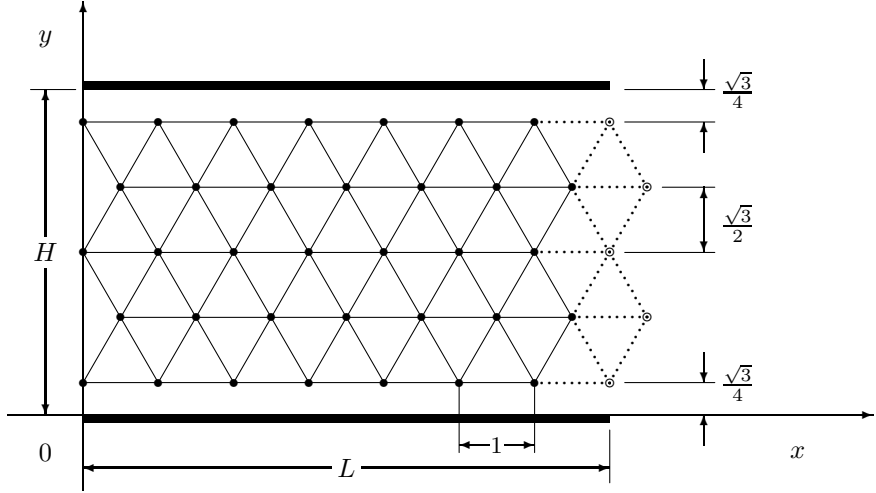

Fig \ref{fig1}. shows the flow domain between two parallel plates and the hexagonal
two-dimensional lattice.
The lattice is periodic in the flow direction, therefore the right nodes,
connected with broken lines, coincide with the left ones. If $N_l $ is the
number of nodes along the $x$ direction, the domain length is $L=N_l $
because the unit length is the triangle side.
The distance between two node rows in the $y$ direction is $\sqrt{3}/2 $.
Because of the "bounce--back" rules imposed in order to satisfy the no-slip
conditions on both planes \cite{b7}, the zero-velocity boundaries are located
at a distance $\sqrt{3}/4 $  from the lowest and highest rows.
If we denote with $N_w $ the number of node rows in the $y$ direction,
the distance between the two real plates is $H=N_w\,\sqrt{3}/2 $.

\subsection{Steady unidirectional flow}

If the local velocity vector has the same direction everywhere and is independent of
distance in the flow direction, the convective rate of change of velocity will
vanish and the acceleration of a fluid element will be
$\partial\vec u /\partial t $, with only one non-zero component.
In addition, if we consider a steady flow, even this acceleration component
will be zero.

The motion between two rest parallel plates can survive the viscous dissipation
of energy only if there is a continuous energy supply to the fluid by a pressure
gradient $\nabla p $, which must be also independent of $x$. When $\nabla p $
is negative, the pressure gradient represents an uniform body force per unit
volume $\rho\vec f $ in the direction of the positive $x$-axis \cite{b8}.

The x component of the equation of motion reduces to:
\begin{equation}
f + \nu\, \frac{\partial^2 u}{\partial y^2} = 0 \label{eq15} 
\end{equation}
The solution which satisfies the boundary conditions $u(0)=0 $ and $u(H)=0 $ is:
\begin{equation}
 u(y)= \frac{f}{\nu}\,\frac{H^2}{2}\,\left(
\frac{y}{H}-\frac{y^2}{H^2} \right) \label{eq16} 
\end{equation}
with the maximum value
\begin{equation}
 u_{max}=0.125\,\frac{\,f\, H^2\,}{\nu} \label{eq17} 
\end{equation}

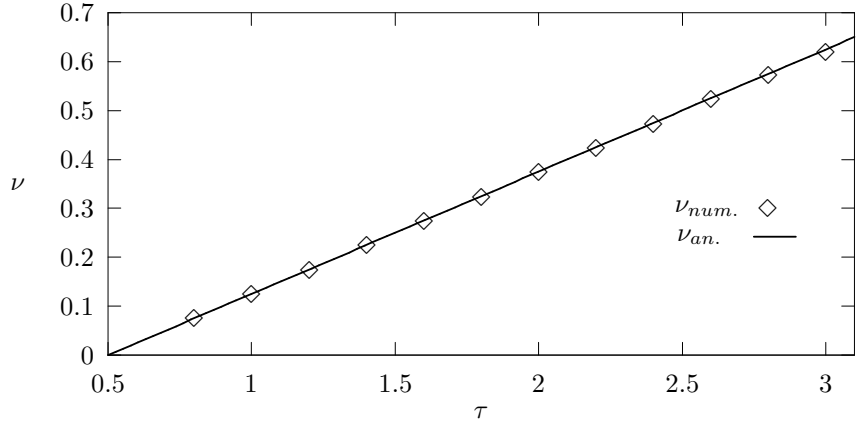
\begin{figure}
\begin{center}
\caption{\label{fig2}
Numerical values for the kinematic viscosity $\nu $, obtained by fitting
the parabolic profiles for $0.8\leq\tau\leq 3 $; the solid line
corresponds to the analytical formula (\protect\ref{eq14}). }
\setlength{\unitlength}{0.240900pt}
\ifx\plotpoint\undefined\newsavebox{\plotpoint}\fi
\sbox{\plotpoint}{\rule[-0.175pt]{0.350pt}{0.350pt}}%
\special{em:linewidth 0.3pt}%
\begin{picture}(1500,809)(100,0)
\tenrm
\put(264,158){\special{em:moveto}}
\put(1436,158){\special{em:lineto}}
\put(264,158){\special{em:moveto}}
\put(284,158){\special{em:lineto}}
\put(1436,158){\special{em:moveto}}
\put(1416,158){\special{em:lineto}}
\put(242,158){\makebox(0,0)[r]{0}}
\put(264,235){\special{em:moveto}}
\put(284,235){\special{em:lineto}}
\put(1436,235){\special{em:moveto}}
\put(1416,235){\special{em:lineto}}
\put(242,235){\makebox(0,0)[r]{0.1}}
\put(264,312){\special{em:moveto}}
\put(284,312){\special{em:lineto}}
\put(1436,312){\special{em:moveto}}
\put(1416,312){\special{em:lineto}}
\put(242,312){\makebox(0,0)[r]{0.2}}
\put(264,389){\special{em:moveto}}
\put(284,389){\special{em:lineto}}
\put(1436,389){\special{em:moveto}}
\put(1416,389){\special{em:lineto}}
\put(242,389){\makebox(0,0)[r]{0.3}}
\put(264,465){\special{em:moveto}}
\put(284,465){\special{em:lineto}}
\put(1436,465){\special{em:moveto}}
\put(1416,465){\special{em:lineto}}
\put(242,465){\makebox(0,0)[r]{0.4}}
\put(264,542){\special{em:moveto}}
\put(284,542){\special{em:lineto}}
\put(1436,542){\special{em:moveto}}
\put(1416,542){\special{em:lineto}}
\put(242,542){\makebox(0,0)[r]{0.5}}
\put(264,619){\special{em:moveto}}
\put(284,619){\special{em:lineto}}
\put(1436,619){\special{em:moveto}}
\put(1416,619){\special{em:lineto}}
\put(242,619){\makebox(0,0)[r]{0.6}}
\put(264,696){\special{em:moveto}}
\put(284,696){\special{em:lineto}}
\put(1436,696){\special{em:moveto}}
\put(1416,696){\special{em:lineto}}
\put(242,696){\makebox(0,0)[r]{0.7}}
\put(264,158){\special{em:moveto}}
\put(264,178){\special{em:lineto}}
\put(264,696){\special{em:moveto}}
\put(264,676){\special{em:lineto}}
\put(264,113){\makebox(0,0){0.5}}
\put(489,158){\special{em:moveto}}
\put(489,178){\special{em:lineto}}
\put(489,696){\special{em:moveto}}
\put(489,676){\special{em:lineto}}
\put(489,113){\makebox(0,0){1}}
\put(715,158){\special{em:moveto}}
\put(715,178){\special{em:lineto}}
\put(715,696){\special{em:moveto}}
\put(715,676){\special{em:lineto}}
\put(715,113){\makebox(0,0){1.5}}
\put(940,158){\special{em:moveto}}
\put(940,178){\special{em:lineto}}
\put(940,696){\special{em:moveto}}
\put(940,676){\special{em:lineto}}
\put(940,113){\makebox(0,0){2}}
\put(1166,158){\special{em:moveto}}
\put(1166,178){\special{em:lineto}}
\put(1166,696){\special{em:moveto}}
\put(1166,676){\special{em:lineto}}
\put(1166,113){\makebox(0,0){2.5}}
\put(1391,158){\special{em:moveto}}
\put(1391,178){\special{em:lineto}}
\put(1391,696){\special{em:moveto}}
\put(1391,676){\special{em:lineto}}
\put(1391,113){\makebox(0,0){3}}
\put(264,158){\special{em:moveto}}
\put(1436,158){\special{em:lineto}}
\put(1436,696){\special{em:lineto}}
\put(264,696){\special{em:lineto}}
\put(264,158){\special{em:lineto}}
\put(111,427){\makebox(0,0)[l]{\shortstack{$ \nu $}}}
\put(850,68){\makebox(0,0){$ \tau $}}
\put(1256,389){\makebox(0,0)[r]{$\nu_{num.}$}}
\put(1300,389){\raisebox{-1.2pt}{\makebox(0,0){$\Diamond$}}}
\put(399,216){\raisebox{-1.2pt}{\makebox(0,0){$\Diamond$}}}
\put(489,254){\raisebox{-1.2pt}{\makebox(0,0){$\Diamond$}}}
\put(580,292){\raisebox{-1.2pt}{\makebox(0,0){$\Diamond$}}}
\put(670,331){\raisebox{-1.2pt}{\makebox(0,0){$\Diamond$}}}
\put(760,369){\raisebox{-1.2pt}{\makebox(0,0){$\Diamond$}}}
\put(850,407){\raisebox{-1.2pt}{\makebox(0,0){$\Diamond$}}}
\put(940,446){\raisebox{-1.2pt}{\makebox(0,0){$\Diamond$}}}
\put(1030,484){\raisebox{-1.2pt}{\makebox(0,0){$\Diamond$}}}
\put(1120,522){\raisebox{-1.2pt}{\makebox(0,0){$\Diamond$}}}
\put(1211,560){\raisebox{-1.2pt}{\makebox(0,0){$\Diamond$}}}
\put(1301,598){\raisebox{-1.2pt}{\makebox(0,0){$\Diamond$}}}
\put(1391,635){\raisebox{-1.2pt}{\makebox(0,0){$\Diamond$}}}
\sbox{\plotpoint}{\rule[-0.350pt]{0.700pt}{0.700pt}}%
\special{em:linewidth 0.7pt}%
\put(1256,344){\makebox(0,0)[r]{$\nu_{an.\mbox{\ \ }}$}}
\put(1278,344){\special{em:moveto}}
\put(1344,344){\special{em:lineto}}
\put(264,158){\special{em:moveto}}
\put(276,163){\special{em:lineto}}
\put(288,168){\special{em:lineto}}
\put(300,173){\special{em:lineto}}
\put(311,178){\special{em:lineto}}
\put(323,183){\special{em:lineto}}
\put(335,188){\special{em:lineto}}
\put(347,193){\special{em:lineto}}
\put(359,198){\special{em:lineto}}
\put(371,203){\special{em:lineto}}
\put(382,208){\special{em:lineto}}
\put(394,214){\special{em:lineto}}
\put(406,219){\special{em:lineto}}
\put(418,224){\special{em:lineto}}
\put(430,229){\special{em:lineto}}
\put(442,234){\special{em:lineto}}
\put(453,239){\special{em:lineto}}
\put(465,244){\special{em:lineto}}
\put(477,249){\special{em:lineto}}
\put(489,254){\special{em:lineto}}
\put(501,259){\special{em:lineto}}
\put(513,264){\special{em:lineto}}
\put(524,269){\special{em:lineto}}
\put(536,274){\special{em:lineto}}
\put(548,279){\special{em:lineto}}
\put(560,284){\special{em:lineto}}
\put(572,289){\special{em:lineto}}
\put(584,294){\special{em:lineto}}
\put(595,299){\special{em:lineto}}
\put(607,304){\special{em:lineto}}
\put(619,309){\special{em:lineto}}
\put(631,314){\special{em:lineto}}
\put(643,319){\special{em:lineto}}
\put(655,325){\special{em:lineto}}
\put(667,330){\special{em:lineto}}
\put(678,335){\special{em:lineto}}
\put(690,340){\special{em:lineto}}
\put(702,345){\special{em:lineto}}
\put(714,350){\special{em:lineto}}
\put(726,355){\special{em:lineto}}
\put(738,360){\special{em:lineto}}
\put(749,365){\special{em:lineto}}
\put(761,370){\special{em:lineto}}
\put(773,375){\special{em:lineto}}
\put(785,380){\special{em:lineto}}
\put(797,385){\special{em:lineto}}
\put(809,390){\special{em:lineto}}
\put(820,395){\special{em:lineto}}
\put(832,400){\special{em:lineto}}
\put(844,405){\special{em:lineto}}
\put(856,410){\special{em:lineto}}
\put(868,415){\special{em:lineto}}
\put(880,420){\special{em:lineto}}
\put(891,425){\special{em:lineto}}
\put(903,430){\special{em:lineto}}
\put(915,436){\special{em:lineto}}
\put(927,441){\special{em:lineto}}
\put(939,446){\special{em:lineto}}
\put(951,451){\special{em:lineto}}
\put(962,456){\special{em:lineto}}
\put(974,461){\special{em:lineto}}
\put(986,466){\special{em:lineto}}
\put(998,471){\special{em:lineto}}
\put(1010,476){\special{em:lineto}}
\put(1022,481){\special{em:lineto}}
\put(1033,486){\special{em:lineto}}
\put(1045,491){\special{em:lineto}}
\put(1057,496){\special{em:lineto}}
\put(1069,501){\special{em:lineto}}
\put(1081,506){\special{em:lineto}}
\put(1093,511){\special{em:lineto}}
\put(1105,516){\special{em:lineto}}
\put(1116,521){\special{em:lineto}}
\put(1128,526){\special{em:lineto}}
\put(1140,531){\special{em:lineto}}
\put(1152,536){\special{em:lineto}}
\put(1164,542){\special{em:lineto}}
\put(1176,547){\special{em:lineto}}
\put(1187,552){\special{em:lineto}}
\put(1199,557){\special{em:lineto}}
\put(1211,562){\special{em:lineto}}
\put(1223,567){\special{em:lineto}}
\put(1235,572){\special{em:lineto}}
\put(1247,577){\special{em:lineto}}
\put(1258,582){\special{em:lineto}}
\put(1270,587){\special{em:lineto}}
\put(1282,592){\special{em:lineto}}
\put(1294,597){\special{em:lineto}}
\put(1306,602){\special{em:lineto}}
\put(1318,607){\special{em:lineto}}
\put(1329,612){\special{em:lineto}}
\put(1341,617){\special{em:lineto}}
\put(1353,622){\special{em:lineto}}
\put(1365,627){\special{em:lineto}}
\put(1377,632){\special{em:lineto}}
\put(1389,637){\special{em:lineto}}
\put(1400,642){\special{em:lineto}}
\put(1412,647){\special{em:lineto}}
\put(1424,653){\special{em:lineto}}
\put(1436,658){\special{em:lineto}}
\end{picture}
\end{center}
\end{figure}

Both the equations (\ref{eq15}) and (\ref{eq17}) 
can be used in order to make a comparison
between analytical (eq.\ref{eq14}) and numerical results. As an example, the kinematic
viscosity $\nu $ can be calculated by using the second derivative of $u(y) $
from (\ref{eq15}) or the maximum velocity $u_{max} $ from (\ref{eq17}).
The first possibility was prefered in this paper because it takes into account
the whole velocity profile shape. If we use the least square method to fit
a parabola between $(y_i,u_i) $ points, $i=1\ldots N_w $, we will have for the
second derivative
\begin{equation}
 \frac{\partial^2 u}{\partial y^2} = -2\,
\frac{\,\sum_i u_i(y_i H - y_i^2)\,}{\sum_i (y_i H - y_i^2)^2} \label{eq18} 
\end{equation}
The body force per unit volume, $\rho\vec f $, must equalize the decay
of the momentum $\rho u $ during one time step
\begin{equation}
 \rho f = \frac{\partial (\rho u)}{\partial t} \label{eq19} 
\end{equation}
Taking into account that $\delta t=1 $ for LBM, and
\begin{equation}
 \delta(\rho u) = \frac{2\,\delta n}{\sqrt{3}/2} \label{eq20} 
\end{equation}
where $\delta n $ is the value to be added to $n_1 $, respectively subtracted
from $n_4 $ at each step, according with (\ref{eq5}) we obtain:
\begin{equation}
 f = \frac{2 \delta n}{n} \label{eq21} 
\end{equation}
Finally, from (\ref{eq15}),(\ref{eq18}) and (\ref{eq21}) the kinematic viscosity is:
\begin{equation}
 \nu = \frac{\delta n}{n}
\frac{\sum_i (y_i H - y_i^2)^2}{\,\sum_i u_i(y_i H - y_i^2)\,} \label{eq22} 
\end{equation}
The numerical results were obtained using a lattice with $N_l=100 $,
$N_w = 101$ ($H=87.4686$). For $\tau \geq 0.8 $ we considered
$\delta n = 10^{-6} $, and the numerical values for $\nu $ are represented
in Fig. \ref{fig2}, together with the analytical line (\ref{eq14}).
The relative error does not exceed 0.8\%.

A special attention was given for $0.5 < \tau <0.8 $, the results being
presented in Table \ref{tab1}.
It should be noticed that $\delta n$ must be decreased when $\tau $,
respectively $\nu $, diminish in order to have only positive $n_i $ values,
for each node and direction.

\begin{table}
\caption{\label{tab1}
Kinematic viscosities ($\nu_{an}$ from 
(\protect\ref{eq14}) and $\nu_{num}$ obtained
by fitting parabolic velocity profiles) for $0.5 < \tau \leq 0.8 $. }
\begin{center}
\begin{tabular}{ccccccc}
$\tau$ & $\delta n$ & steps &  $u_{max}$ & 
$\nu_{an}$ & $\nu_{num}$ & $ \varepsilon_{\nu}$ [\%] \\
\hline
$ 0.51 $ & $ 10^{-15} $ & $ 2.0\cdot 10^6 $ & $ 3.6\cdot 10^{-10} $ & $ 2.50 \cdot 10^{-3} $ & $ 2.5074\cdot 10^{-3} $ & $ 0.29 $ \\
$ 0.52 $ & $ 10^{-14} $ & $ 8.0\cdot 10^5 $ & $ 1.8\cdot 10^{-9}  $ & $ 5.00 \cdot 10^{-3} $ & $ 5.0353\cdot 10^{-3} $ & $ 0.71 $ \\
$ 0.53 $ & $ 10^{-13} $ & $ 6.0\cdot 10^5 $ & $ 1.2\cdot 10^{-8}  $ & $ 7.50 \cdot 10^{-3} $ & $ 7.5255\cdot 10^{-3} $ & $ 0.34 $ \\
$ 0.54 $ & $ 10^{-12} $ & $ 4.0\cdot 10^5 $ & $ 9.1\cdot 10^{-8}  $ & $ 1.00 \cdot 10^{-2} $ & $ 1.0059\cdot 10^{-2} $ & $ 0.59 $ \\
$ 0.55 $ & $ 10^{-12} $ & $ 3.5\cdot 10^5 $ & $ 7.3\cdot 10^{-8}  $ & $ 1.25 \cdot 10^{-2} $ & $ 1.2546\cdot 10^{-2} $ & $ 0.37 $ \\
$ 0.56 $ & $ 10^{-11} $ & $ 3.5\cdot 10^5 $ & $ 6.1\cdot 10^{-7}  $ & $ 1.50 \cdot 10^{-2} $ & $ 1.5019\cdot 10^{-2} $ & $ 0.13 $ \\
$ 0.57 $ & $ 10^{-11} $ & $ 3.0\cdot 10^5 $ & $ 5.2\cdot 10^{-7}  $ & $ 1.75 \cdot 10^{-2} $ & $ 1.7522\cdot 10^{-2} $ & $ 0.13 $ \\
$ 0.58 $ & $ 10^{-11} $ & $ 3.0\cdot 10^5 $ & $ 4.6\cdot 10^{-7}  $ & $ 2.00 \cdot 10^{-2} $ & $ 2.0011\cdot 10^{-2} $ & $ 0.06 $ \\
$ 0.59 $ & $ 10^{-10} $ & $ 3.0\cdot 10^5 $ & $ 4.0\cdot 10^{-6}  $ & $ 2.25 \cdot 10^{-2} $ & $ 2.2506\cdot 10^{-2} $ & $ 0.03 $ \\
$ 0.60 $ & $ 10^{-10} $ & $ 3.0\cdot 10^5 $ & $ 3.6\cdot 10^{-6}  $ & $ 2.50 \cdot 10^{-2} $ & $ 2.5004\cdot 10^{-2} $ & $ 0.02 $ \\
$ 0.62 $ & $ 10^{-9}  $ & $ 2.5\cdot 10^5 $ & $ 3.0\cdot 10^{-5}  $ & $ 3.00 \cdot 10^{-2} $ & $ 3.0005\cdot 10^{-2} $ & $ 0.02 $ \\
$ 0.64 $ & $ 10^{-9}  $ & $ 2.0\cdot 10^5 $ & $ 2.6\cdot 10^{-5}  $ & $ 3.50 \cdot 10^{-2} $ & $ 3.5008\cdot 10^{-2} $ & $ 0.02 $ \\
$ 0.66 $ & $ 10^{-9}  $ & $ 2.0\cdot 10^5 $ & $ 2.3\cdot 10^{-5}  $ & $ 4.00 \cdot 10^{-2} $ & $ 4.0005\cdot 10^{-2} $ & $ 0.01 $ \\
$ 0.68 $ & $ 10^{-8}  $ & $ 1.0\cdot 10^5 $ & $ 2.0\cdot 10^{-4}  $ & $ 4.50 \cdot 10^{-2} $ & $ 4.5140\cdot 10^{-2} $ & $ 0.31 $ \\
$ 0.70 $ & $ 10^{-8}  $ & $ 1.0\cdot 10^5 $ & $ 1.8\cdot 10^{-4}  $ & $ 5.00 \cdot 10^{-2} $ & $ 5.0083\cdot 10^{-2} $ & $ 0.17 $ \\
$ 0.72 $ & $ 10^{-8}  $ & $ 1.0\cdot 10^5 $ & $ 1.7\cdot 10^{-4}  $ & $ 5.50 \cdot 10^{-2} $ & $ 5.5050\cdot 10^{-2} $ & $ 0.09 $ \\
$ 0.74 $ & $ 10^{-7}  $ & $ 1.0\cdot 10^5 $ & $ 1.5\cdot 10^{-3}  $ & $ 6.00 \cdot 10^{-2} $ & $ 6.0030\cdot 10^{-2} $ & $ 0.05 $ \\
$ 0.76 $ & $ 10^{-7}  $ & $ 1.0\cdot 10^5 $ & $ 1.4\cdot 10^{-3}  $ & $ 6.50 \cdot 10^{-2} $ & $ 6.5018\cdot 10^{-2} $ & $ 0.03 $ \\
$ 0.78 $ & $ 10^{-7}  $ & $ 1.0\cdot 10^5 $ & $ 1.3\cdot 10^{-3}  $ & $ 7.00 \cdot 10^{-2} $ & $ 7.0012\cdot 10^{-2} $ & $ 0.02 $ \\
$ 0.80 $ & $ 10^{-6}  $ & $ 5.0\cdot 10^4 $ & $ 1.2\cdot 10^{-2}  $ & $ 7.50 \cdot 10^{-2} $ & $ 7.5601\cdot 10^{-2} $ & $ 0.80 $ \\
\end{tabular}
\end{center}
\end{table}

Table \ref{tab1} also shows the number of steps needed to reach the steady state and
the corresponding maximum velocity $u_{max} $ (see eq.\ref{eq17}). The relative
difference between theoretical and analytical values,
$\varepsilon_{\nu}=\left|\nu_{an}-\nu_{num}\right|/\nu_{an}$,
does not exceed one percent.

\begin{figure}
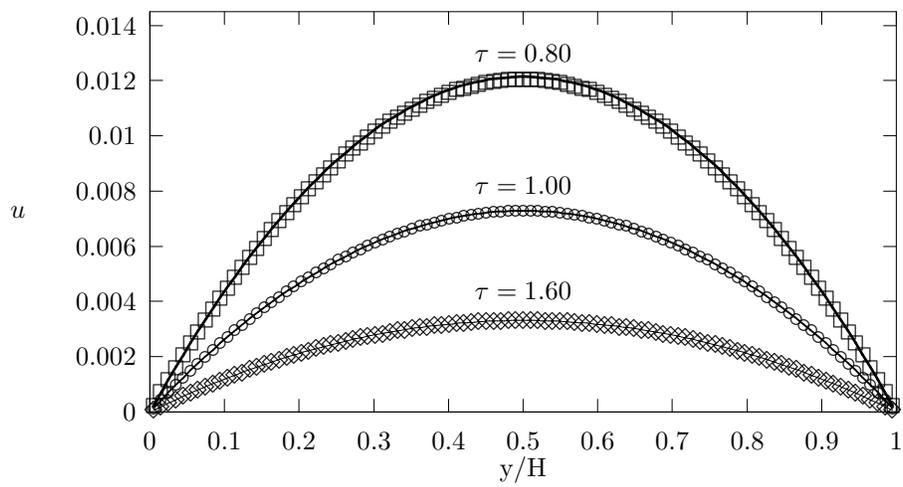

\begin{center}
\caption{\label{fig3} 
Parabolic velocity profiles between parallel rest plates,
for $\tau = $ 0.80, 1.00 and 1.60.}
\setlength{\unitlength}{0.240900pt}
\ifx\plotpoint\undefined\newsavebox{\plotpoint}\fi
\sbox{\plotpoint}{\rule[-0.175pt]{0.350pt}{0.350pt}}%
\special{em:linewidth 0.3pt}%

\end{center}
\end{figure}

Three velocity profiles are represented in Fig.\ref{fig3}. A very good agreement
exist between the theoretical parabola (solid curve) and the numerical
results.

\subsection{Entrance--region steady flow}

If we consider wind-tunnel conditions at $x=0 $, i.e. an uniform velocity $U$
in the inlet section, the velocity profile must become parabolic far downstream
\begin{equation}
u(y) = 6\, U \left( \frac{y}{H} - \frac{y^2}{H^2} \right) \label{eq23}
\end{equation}

An analytical approximate description of such experimental situation has been
performed by Schlichting \cite{b9}. At the beginning, i.e. at small distances
from the inlet section, the boundary layers will grow in the same way as along
a flat plate at zero incidence. The resulting velocity profile will consist
of two boundary-layer profiles on the two walls joined in the center by a line
of constant velocity. Since the flow rate must be the same for every section, the
decrease in the rate of flow near the walls which is due to friction must be
compensated by a corresponding increase near the axis. Thus the boundary layer
is formed under the influence of an accelerated external flow, as distinct from
the case of the flat plate. At larger distances from the inlet section the two
boundary layers gradually merge into each other , and finally the velocity
profile is transformed  asymptotically into the parabolic 
distribution (\ref{eq23}).
Schlichting estimated the inlet length as $l_E = 0.04\, H\, Re $, where $Re $
denotes the Reynolds number referred to the width of the channel, $U\, H/\nu $.
A comparison of results obtained with Lattice Gas Method to those computed by
Schlichling was presented in \cite{b10}.

The present LBM simulation was performed using a lattice
with $N_w=210 $, therefore $H=201\, \sqrt{3}/2 = 174.04 $. The average
velocity between plates was chosen $U = 0.01 $ and
the kinematic viscosity $\nu = 0.125 $ (for $\tau = 1.0 $). It results
a small Reynolds number $Re = 13.925 $ and $l_E = 97 $. In this case,
the Schlichting solution is not valid because the boundary layer approximation
cannot be used and it is expected that the $l_E $
value becomes larger. To overcome this situation, we choose $L=200 $.

\begin{figure}
\caption{\label{fig4} Velocity profiles at different
distances $x/H = $ 0.0718~(a),
0.1436~(b), 0.2873~(c), 0.5746~(d) from the inlet section
(solid curves -- LBM results, dots -- FEM solution).  }
\setlength{\unitlength}{0.240900pt}
\ifx\plotpoint\undefined\newsavebox{\plotpoint}\fi
\sbox{\plotpoint}{\rule[-0.175pt]{0.350pt}{0.350pt}}%
\special{em:linewidth 0.3pt}%
\begin{picture}(550,629)(0,0)
\tenrm
\put(264,158){\special{em:moveto}}
\put(686,158){\special{em:lineto}}
\put(264,158){\special{em:moveto}}
\put(264,516){\special{em:lineto}}
\put(264,158){\special{em:moveto}}
\put(284,158){\special{em:lineto}}
\put(686,158){\special{em:moveto}}
\put(666,158){\special{em:lineto}}
\put(242,158){\makebox(0,0)[r]{0}}
\put(264,230){\special{em:moveto}}
\put(284,230){\special{em:lineto}}
\put(686,230){\special{em:moveto}}
\put(666,230){\special{em:lineto}}
\put(242,230){\makebox(0,0)[r]{0.1}}
\put(264,301){\special{em:moveto}}
\put(284,301){\special{em:lineto}}
\put(686,301){\special{em:moveto}}
\put(666,301){\special{em:lineto}}
\put(242,301){\makebox(0,0)[r]{0.2}}
\put(264,373){\special{em:moveto}}
\put(284,373){\special{em:lineto}}
\put(686,373){\special{em:moveto}}
\put(666,373){\special{em:lineto}}
\put(242,373){\makebox(0,0)[r]{0.3}}
\put(264,444){\special{em:moveto}}
\put(284,444){\special{em:lineto}}
\put(686,444){\special{em:moveto}}
\put(666,444){\special{em:lineto}}
\put(242,444){\makebox(0,0)[r]{0.4}}
\put(264,516){\special{em:moveto}}
\put(284,516){\special{em:lineto}}
\put(686,516){\special{em:moveto}}
\put(666,516){\special{em:lineto}}
\put(242,516){\makebox(0,0)[r]{0.5}}
\put(264,158){\special{em:moveto}}
\put(264,178){\special{em:lineto}}
\put(264,516){\special{em:moveto}}
\put(264,496){\special{em:lineto}}
\put(264,113){\makebox(0,0){0}}
\put(405,158){\special{em:moveto}}
\put(405,178){\special{em:lineto}}
\put(405,516){\special{em:moveto}}
\put(405,496){\special{em:lineto}}
\put(405,113){\makebox(0,0){0.5}}
\put(545,158){\special{em:moveto}}
\put(545,178){\special{em:lineto}}
\put(545,516){\special{em:moveto}}
\put(545,496){\special{em:lineto}}
\put(545,113){\makebox(0,0){1}}
\put(686,158){\special{em:moveto}}
\put(686,178){\special{em:lineto}}
\put(686,516){\special{em:moveto}}
\put(686,496){\special{em:lineto}}
\put(686,113){\makebox(0,0){1.5}}
\put(264,158){\special{em:moveto}}
\put(686,158){\special{em:lineto}}
\put(686,516){\special{em:lineto}}
\put(264,516){\special{em:lineto}}
\put(264,158){\special{em:lineto}}
\put(133,337){\makebox(0,0)[l]{\shortstack{$\frac{\,y\,}{H}$}}}
\put(475,68){\makebox(0,0){$u/U$}}
\put(337,337){\makebox(0,0){\framebox{{\bf a}}}}
\put(377,459){\makebox(0,0)[r]{{\scriptsize LBM}}}
\put(399,459){\special{em:moveto}}
\put(465,459){\special{em:lineto}}
\put(264,158){\special{em:moveto}}
\put(324,163){\special{em:lineto}}
\put(394,170){\special{em:lineto}}
\put(449,178){\special{em:lineto}}
\put(490,185){\special{em:lineto}}
\put(520,192){\special{em:lineto}}
\put(540,199){\special{em:lineto}}
\put(553,206){\special{em:lineto}}
\put(561,213){\special{em:lineto}}
\put(565,220){\special{em:lineto}}
\put(568,227){\special{em:lineto}}
\put(569,235){\special{em:lineto}}
\put(569,242){\special{em:lineto}}
\put(569,249){\special{em:lineto}}
\put(568,256){\special{em:lineto}}
\put(567,263){\special{em:lineto}}
\put(566,270){\special{em:lineto}}
\put(565,277){\special{em:lineto}}
\put(564,284){\special{em:lineto}}
\put(563,292){\special{em:lineto}}
\put(562,299){\special{em:lineto}}
\put(561,306){\special{em:lineto}}
\put(560,313){\special{em:lineto}}
\put(559,320){\special{em:lineto}}
\put(559,327){\special{em:lineto}}
\put(558,334){\special{em:lineto}}
\put(558,341){\special{em:lineto}}
\put(557,349){\special{em:lineto}}
\put(557,356){\special{em:lineto}}
\put(556,363){\special{em:lineto}}
\put(556,370){\special{em:lineto}}
\put(555,377){\special{em:lineto}}
\put(555,384){\special{em:lineto}}
\put(555,391){\special{em:lineto}}
\put(554,398){\special{em:lineto}}
\put(554,406){\special{em:lineto}}
\put(554,413){\special{em:lineto}}
\put(554,420){\special{em:lineto}}
\put(553,427){\special{em:lineto}}
\put(553,434){\special{em:lineto}}
\put(553,441){\special{em:lineto}}
\put(553,448){\special{em:lineto}}
\put(553,455){\special{em:lineto}}
\put(553,463){\special{em:lineto}}
\put(553,470){\special{em:lineto}}
\put(553,477){\special{em:lineto}}
\put(553,484){\special{em:lineto}}
\put(552,491){\special{em:lineto}}
\put(552,498){\special{em:lineto}}
\put(552,505){\special{em:lineto}}
\put(552,512){\special{em:lineto}}
\put(377,414){\makebox(0,0)[r]{{\scriptsize FEM}}}
\put(421,414){\circle{12}}
\put(264,158){\circle{12}}
\put(303,160){\circle{12}}
\put(354,167){\circle{12}}
\put(424,175){\circle{12}}
\put(495,183){\circle{12}}
\put(534,192){\circle{12}}
\put(559,202){\circle{12}}
\put(571,213){\circle{12}}
\put(576,225){\circle{12}}
\put(573,238){\circle{12}}
\put(571,251){\circle{12}}
\put(568,267){\circle{12}}
\put(565,283){\circle{12}}
\put(562,301){\circle{12}}
\put(559,321){\circle{12}}
\put(557,342){\circle{12}}
\put(554,365){\circle{12}}
\put(551,391){\circle{12}}
\put(551,418){\circle{12}}
\put(551,448){\circle{12}}
\put(551,481){\circle{12}}
\put(548,516){\circle{12}}
\end{picture}
\hspace{2em}
\setlength{\unitlength}{0.240900pt}
\ifx\plotpoint\undefined\newsavebox{\plotpoint}\fi
\sbox{\plotpoint}{\rule[-0.175pt]{0.350pt}{0.350pt}}%
\special{em:linewidth 0.3pt}%
\begin{picture}(550,629)(0,0)
\tenrm
\put(264,158){\special{em:moveto}}
\put(686,158){\special{em:lineto}}
\put(264,158){\special{em:moveto}}
\put(264,516){\special{em:lineto}}
\put(264,158){\special{em:moveto}}
\put(284,158){\special{em:lineto}}
\put(686,158){\special{em:moveto}}
\put(666,158){\special{em:lineto}}
\put(242,158){\makebox(0,0)[r]{0}}
\put(264,230){\special{em:moveto}}
\put(284,230){\special{em:lineto}}
\put(686,230){\special{em:moveto}}
\put(666,230){\special{em:lineto}}
\put(242,230){\makebox(0,0)[r]{0.1}}
\put(264,301){\special{em:moveto}}
\put(284,301){\special{em:lineto}}
\put(686,301){\special{em:moveto}}
\put(666,301){\special{em:lineto}}
\put(242,301){\makebox(0,0)[r]{0.2}}
\put(264,373){\special{em:moveto}}
\put(284,373){\special{em:lineto}}
\put(686,373){\special{em:moveto}}
\put(666,373){\special{em:lineto}}
\put(242,373){\makebox(0,0)[r]{0.3}}
\put(264,444){\special{em:moveto}}
\put(284,444){\special{em:lineto}}
\put(686,444){\special{em:moveto}}
\put(666,444){\special{em:lineto}}
\put(242,444){\makebox(0,0)[r]{0.4}}
\put(264,516){\special{em:moveto}}
\put(284,516){\special{em:lineto}}
\put(686,516){\special{em:moveto}}
\put(666,516){\special{em:lineto}}
\put(242,516){\makebox(0,0)[r]{0.5}}
\put(264,158){\special{em:moveto}}
\put(264,178){\special{em:lineto}}
\put(264,516){\special{em:moveto}}
\put(264,496){\special{em:lineto}}
\put(264,113){\makebox(0,0){0}}
\put(405,158){\special{em:moveto}}
\put(405,178){\special{em:lineto}}
\put(405,516){\special{em:moveto}}
\put(405,496){\special{em:lineto}}
\put(405,113){\makebox(0,0){0.5}}
\put(545,158){\special{em:moveto}}
\put(545,178){\special{em:lineto}}
\put(545,516){\special{em:moveto}}
\put(545,496){\special{em:lineto}}
\put(545,113){\makebox(0,0){1}}
\put(686,158){\special{em:moveto}}
\put(686,178){\special{em:lineto}}
\put(686,516){\special{em:moveto}}
\put(686,496){\special{em:lineto}}
\put(686,113){\makebox(0,0){1.5}}
\put(264,158){\special{em:moveto}}
\put(686,158){\special{em:lineto}}
\put(686,516){\special{em:lineto}}
\put(264,516){\special{em:lineto}}
\put(264,158){\special{em:lineto}}
\put(133,337){\makebox(0,0)[l]{\shortstack{$\frac{\,y\,}{H}$}}}
\put(475,68){\makebox(0,0){$u/U$}}
\put(337,337){\makebox(0,0){\framebox{{\bf b}}}}
\put(377,459){\makebox(0,0)[r]{{\scriptsize LBM}}}
\put(399,459){\special{em:moveto}}
\put(465,459){\special{em:lineto}}
\put(275,160){\special{em:moveto}}
\put(318,167){\special{em:lineto}}
\put(359,174){\special{em:lineto}}
\put(396,181){\special{em:lineto}}
\put(429,188){\special{em:lineto}}
\put(458,195){\special{em:lineto}}
\put(483,203){\special{em:lineto}}
\put(504,210){\special{em:lineto}}
\put(522,217){\special{em:lineto}}
\put(537,224){\special{em:lineto}}
\put(548,231){\special{em:lineto}}
\put(558,238){\special{em:lineto}}
\put(565,245){\special{em:lineto}}
\put(571,252){\special{em:lineto}}
\put(575,260){\special{em:lineto}}
\put(578,267){\special{em:lineto}}
\put(580,274){\special{em:lineto}}
\put(581,281){\special{em:lineto}}
\put(582,288){\special{em:lineto}}
\put(582,295){\special{em:lineto}}
\put(582,302){\special{em:lineto}}
\put(581,309){\special{em:lineto}}
\put(580,317){\special{em:lineto}}
\put(580,324){\special{em:lineto}}
\put(579,331){\special{em:lineto}}
\put(578,338){\special{em:lineto}}
\put(577,345){\special{em:lineto}}
\put(576,352){\special{em:lineto}}
\put(575,359){\special{em:lineto}}
\put(574,366){\special{em:lineto}}
\put(573,374){\special{em:lineto}}
\put(572,381){\special{em:lineto}}
\put(571,388){\special{em:lineto}}
\put(570,395){\special{em:lineto}}
\put(570,402){\special{em:lineto}}
\put(569,409){\special{em:lineto}}
\put(568,416){\special{em:lineto}}
\put(568,423){\special{em:lineto}}
\put(567,431){\special{em:lineto}}
\put(567,438){\special{em:lineto}}
\put(566,445){\special{em:lineto}}
\put(566,452){\special{em:lineto}}
\put(565,459){\special{em:lineto}}
\put(565,466){\special{em:lineto}}
\put(565,473){\special{em:lineto}}
\put(564,480){\special{em:lineto}}
\put(564,488){\special{em:lineto}}
\put(564,495){\special{em:lineto}}
\put(564,502){\special{em:lineto}}
\put(564,509){\special{em:lineto}}
\put(564,516){\special{em:lineto}}
\put(377,414){\makebox(0,0)[r]{{\scriptsize FEM}}}
\put(421,414){\circle{12}}
\put(264,158){\circle{12}}
\put(270,160){\circle{12}}
\put(323,167){\circle{12}}
\put(362,175){\circle{12}}
\put(407,183){\circle{12}}
\put(450,192){\circle{12}}
\put(486,202){\circle{12}}
\put(520,213){\circle{12}}
\put(545,225){\circle{12}}
\put(562,238){\circle{12}}
\put(576,251){\circle{12}}
\put(582,267){\circle{12}}
\put(585,283){\circle{12}}
\put(585,301){\circle{12}}
\put(582,321){\circle{12}}
\put(576,342){\circle{12}}
\put(573,365){\circle{12}}
\put(571,391){\circle{12}}
\put(565,418){\circle{12}}
\put(565,448){\circle{12}}
\put(562,481){\circle{12}}
\put(562,516){\circle{12}}
\end{picture}
\\[-2em]

\setlength{\unitlength}{0.240900pt}
\ifx\plotpoint\undefined\newsavebox{\plotpoint}\fi
\sbox{\plotpoint}{\rule[-0.175pt]{0.350pt}{0.350pt}}%
\special{em:linewidth 0.3pt}%
\begin{picture}(550,629)(0,0)
\tenrm
\put(264,158){\special{em:moveto}}
\put(686,158){\special{em:lineto}}
\put(264,158){\special{em:moveto}}
\put(264,516){\special{em:lineto}}
\put(264,158){\special{em:moveto}}
\put(284,158){\special{em:lineto}}
\put(686,158){\special{em:moveto}}
\put(666,158){\special{em:lineto}}
\put(242,158){\makebox(0,0)[r]{0}}
\put(264,230){\special{em:moveto}}
\put(284,230){\special{em:lineto}}
\put(686,230){\special{em:moveto}}
\put(666,230){\special{em:lineto}}
\put(242,230){\makebox(0,0)[r]{0.1}}
\put(264,301){\special{em:moveto}}
\put(284,301){\special{em:lineto}}
\put(686,301){\special{em:moveto}}
\put(666,301){\special{em:lineto}}
\put(242,301){\makebox(0,0)[r]{0.2}}
\put(264,373){\special{em:moveto}}
\put(284,373){\special{em:lineto}}
\put(686,373){\special{em:moveto}}
\put(666,373){\special{em:lineto}}
\put(242,373){\makebox(0,0)[r]{0.3}}
\put(264,444){\special{em:moveto}}
\put(284,444){\special{em:lineto}}
\put(686,444){\special{em:moveto}}
\put(666,444){\special{em:lineto}}
\put(242,444){\makebox(0,0)[r]{0.4}}
\put(264,516){\special{em:moveto}}
\put(284,516){\special{em:lineto}}
\put(686,516){\special{em:moveto}}
\put(666,516){\special{em:lineto}}
\put(242,516){\makebox(0,0)[r]{0.5}}
\put(264,158){\special{em:moveto}}
\put(264,178){\special{em:lineto}}
\put(264,516){\special{em:moveto}}
\put(264,496){\special{em:lineto}}
\put(264,113){\makebox(0,0){0}}
\put(405,158){\special{em:moveto}}
\put(405,178){\special{em:lineto}}
\put(405,516){\special{em:moveto}}
\put(405,496){\special{em:lineto}}
\put(405,113){\makebox(0,0){0.5}}
\put(545,158){\special{em:moveto}}
\put(545,178){\special{em:lineto}}
\put(545,516){\special{em:moveto}}
\put(545,496){\special{em:lineto}}
\put(545,113){\makebox(0,0){1}}
\put(686,158){\special{em:moveto}}
\put(686,178){\special{em:lineto}}
\put(686,516){\special{em:moveto}}
\put(686,496){\special{em:lineto}}
\put(686,113){\makebox(0,0){1.5}}
\put(264,158){\special{em:moveto}}
\put(686,158){\special{em:lineto}}
\put(686,516){\special{em:lineto}}
\put(264,516){\special{em:lineto}}
\put(264,158){\special{em:lineto}}
\put(133,337){\makebox(0,0)[l]{\shortstack{$\frac{\,y\,}{H}$}}}
\put(475,68){\makebox(0,0){$u/U$}}
\put(337,337){\makebox(0,0){\framebox{{\bf c}}}}
\put(377,459){\makebox(0,0)[r]{{\scriptsize LBM}}}
\put(399,459){\special{em:moveto}}
\put(465,459){\special{em:lineto}}
\put(270,160){\special{em:moveto}}
\put(296,167){\special{em:lineto}}
\put(321,174){\special{em:lineto}}
\put(345,181){\special{em:lineto}}
\put(368,188){\special{em:lineto}}
\put(390,195){\special{em:lineto}}
\put(411,203){\special{em:lineto}}
\put(430,210){\special{em:lineto}}
\put(449,217){\special{em:lineto}}
\put(466,224){\special{em:lineto}}
\put(482,231){\special{em:lineto}}
\put(497,238){\special{em:lineto}}
\put(511,245){\special{em:lineto}}
\put(524,252){\special{em:lineto}}
\put(535,260){\special{em:lineto}}
\put(545,267){\special{em:lineto}}
\put(554,274){\special{em:lineto}}
\put(562,281){\special{em:lineto}}
\put(570,288){\special{em:lineto}}
\put(576,295){\special{em:lineto}}
\put(581,302){\special{em:lineto}}
\put(586,309){\special{em:lineto}}
\put(590,317){\special{em:lineto}}
\put(593,324){\special{em:lineto}}
\put(596,331){\special{em:lineto}}
\put(598,338){\special{em:lineto}}
\put(600,345){\special{em:lineto}}
\put(602,352){\special{em:lineto}}
\put(603,359){\special{em:lineto}}
\put(604,366){\special{em:lineto}}
\put(604,374){\special{em:lineto}}
\put(604,381){\special{em:lineto}}
\put(605,388){\special{em:lineto}}
\put(605,395){\special{em:lineto}}
\put(604,402){\special{em:lineto}}
\put(604,409){\special{em:lineto}}
\put(604,416){\special{em:lineto}}
\put(604,423){\special{em:lineto}}
\put(603,431){\special{em:lineto}}
\put(603,438){\special{em:lineto}}
\put(602,445){\special{em:lineto}}
\put(602,452){\special{em:lineto}}
\put(602,459){\special{em:lineto}}
\put(601,466){\special{em:lineto}}
\put(601,473){\special{em:lineto}}
\put(601,480){\special{em:lineto}}
\put(600,488){\special{em:lineto}}
\put(600,495){\special{em:lineto}}
\put(600,502){\special{em:lineto}}
\put(600,509){\special{em:lineto}}
\put(600,516){\special{em:lineto}}
\put(377,414){\makebox(0,0)[r]{{\scriptsize FEM}}}
\put(421,414){\circle{12}}
\put(264,158){\circle{12}}
\put(270,160){\circle{12}}
\put(298,167){\circle{12}}
\put(323,175){\circle{12}}
\put(351,183){\circle{12}}
\put(379,192){\circle{12}}
\put(410,202){\circle{12}}
\put(441,213){\circle{12}}
\put(469,225){\circle{12}}
\put(498,238){\circle{12}}
\put(526,251){\circle{12}}
\put(548,267){\circle{12}}
\put(568,283){\circle{12}}
\put(582,301){\circle{12}}
\put(593,321){\circle{12}}
\put(602,342){\circle{12}}
\put(604,365){\circle{12}}
\put(604,391){\circle{12}}
\put(604,418){\circle{12}}
\put(602,448){\circle{12}}
\put(599,481){\circle{12}}
\put(599,516){\circle{12}}
\end{picture}
\hspace{2em}
\setlength{\unitlength}{0.240900pt}
\ifx\plotpoint\undefined\newsavebox{\plotpoint}\fi
\sbox{\plotpoint}{\rule[-0.175pt]{0.350pt}{0.350pt}}%
\special{em:linewidth 0.3pt}%
\begin{picture}(550,629)(0,0)
\tenrm
\put(264,158){\special{em:moveto}}
\put(686,158){\special{em:lineto}}
\put(264,158){\special{em:moveto}}
\put(264,516){\special{em:lineto}}
\put(264,158){\special{em:moveto}}
\put(284,158){\special{em:lineto}}
\put(686,158){\special{em:moveto}}
\put(666,158){\special{em:lineto}}
\put(242,158){\makebox(0,0)[r]{0}}
\put(264,230){\special{em:moveto}}
\put(284,230){\special{em:lineto}}
\put(686,230){\special{em:moveto}}
\put(666,230){\special{em:lineto}}
\put(242,230){\makebox(0,0)[r]{0.1}}
\put(264,301){\special{em:moveto}}
\put(284,301){\special{em:lineto}}
\put(686,301){\special{em:moveto}}
\put(666,301){\special{em:lineto}}
\put(242,301){\makebox(0,0)[r]{0.2}}
\put(264,373){\special{em:moveto}}
\put(284,373){\special{em:lineto}}
\put(686,373){\special{em:moveto}}
\put(666,373){\special{em:lineto}}
\put(242,373){\makebox(0,0)[r]{0.3}}
\put(264,444){\special{em:moveto}}
\put(284,444){\special{em:lineto}}
\put(686,444){\special{em:moveto}}
\put(666,444){\special{em:lineto}}
\put(242,444){\makebox(0,0)[r]{0.4}}
\put(264,516){\special{em:moveto}}
\put(284,516){\special{em:lineto}}
\put(686,516){\special{em:moveto}}
\put(666,516){\special{em:lineto}}
\put(242,516){\makebox(0,0)[r]{0.5}}
\put(264,158){\special{em:moveto}}
\put(264,178){\special{em:lineto}}
\put(264,516){\special{em:moveto}}
\put(264,496){\special{em:lineto}}
\put(264,113){\makebox(0,0){0}}
\put(405,158){\special{em:moveto}}
\put(405,178){\special{em:lineto}}
\put(405,516){\special{em:moveto}}
\put(405,496){\special{em:lineto}}
\put(405,113){\makebox(0,0){0.5}}
\put(545,158){\special{em:moveto}}
\put(545,178){\special{em:lineto}}
\put(545,516){\special{em:moveto}}
\put(545,496){\special{em:lineto}}
\put(545,113){\makebox(0,0){1}}
\put(686,158){\special{em:moveto}}
\put(686,178){\special{em:lineto}}
\put(686,516){\special{em:moveto}}
\put(686,496){\special{em:lineto}}
\put(686,113){\makebox(0,0){1.5}}
\put(264,158){\special{em:moveto}}
\put(686,158){\special{em:lineto}}
\put(686,516){\special{em:lineto}}
\put(264,516){\special{em:lineto}}
\put(264,158){\special{em:lineto}}
\put(133,337){\makebox(0,0)[l]{\shortstack{$\frac{\,y\,}{H}$}}}
\put(475,68){\makebox(0,0){$u/U$}}
\put(337,337){\makebox(0,0){\framebox{{\bf d}}}}
\put(377,459){\makebox(0,0)[r]{{\scriptsize LBM}}}
\put(399,459){\special{em:moveto}}
\put(465,459){\special{em:lineto}}
\put(269,160){\special{em:moveto}}
\put(287,167){\special{em:lineto}}
\put(304,174){\special{em:lineto}}
\put(322,181){\special{em:lineto}}
\put(338,188){\special{em:lineto}}
\put(355,195){\special{em:lineto}}
\put(371,203){\special{em:lineto}}
\put(387,210){\special{em:lineto}}
\put(402,217){\special{em:lineto}}
\put(417,224){\special{em:lineto}}
\put(431,231){\special{em:lineto}}
\put(445,238){\special{em:lineto}}
\put(459,245){\special{em:lineto}}
\put(472,252){\special{em:lineto}}
\put(484,260){\special{em:lineto}}
\put(496,267){\special{em:lineto}}
\put(508,274){\special{em:lineto}}
\put(519,281){\special{em:lineto}}
\put(529,288){\special{em:lineto}}
\put(539,295){\special{em:lineto}}
\put(549,302){\special{em:lineto}}
\put(558,309){\special{em:lineto}}
\put(566,317){\special{em:lineto}}
\put(574,324){\special{em:lineto}}
\put(582,331){\special{em:lineto}}
\put(589,338){\special{em:lineto}}
\put(596,345){\special{em:lineto}}
\put(602,352){\special{em:lineto}}
\put(608,359){\special{em:lineto}}
\put(613,366){\special{em:lineto}}
\put(618,374){\special{em:lineto}}
\put(623,381){\special{em:lineto}}
\put(627,388){\special{em:lineto}}
\put(631,395){\special{em:lineto}}
\put(635,402){\special{em:lineto}}
\put(638,409){\special{em:lineto}}
\put(641,416){\special{em:lineto}}
\put(644,423){\special{em:lineto}}
\put(646,431){\special{em:lineto}}
\put(648,438){\special{em:lineto}}
\put(650,445){\special{em:lineto}}
\put(652,452){\special{em:lineto}}
\put(654,459){\special{em:lineto}}
\put(655,466){\special{em:lineto}}
\put(656,473){\special{em:lineto}}
\put(657,480){\special{em:lineto}}
\put(658,488){\special{em:lineto}}
\put(658,495){\special{em:lineto}}
\put(659,502){\special{em:lineto}}
\put(659,509){\special{em:lineto}}
\put(659,516){\special{em:lineto}}
\put(377,414){\makebox(0,0)[r]{{\scriptsize FEM}}}
\put(421,414){\circle{12}}
\put(264,158){\circle{12}}
\put(270,160){\circle{12}}
\put(287,167){\circle{12}}
\put(306,175){\circle{12}}
\put(326,183){\circle{12}}
\put(348,192){\circle{12}}
\put(371,202){\circle{12}}
\put(393,213){\circle{12}}
\put(419,225){\circle{12}}
\put(444,238){\circle{12}}
\put(469,251){\circle{12}}
\put(498,267){\circle{12}}
\put(523,283){\circle{12}}
\put(548,301){\circle{12}}
\put(573,321){\circle{12}}
\put(593,342){\circle{12}}
\put(613,365){\circle{12}}
\put(630,391){\circle{12}}
\put(644,418){\circle{12}}
\put(652,448){\circle{12}}
\put(658,481){\circle{12}}
\put(661,516){\circle{12}}
\end{picture}
\end{figure}
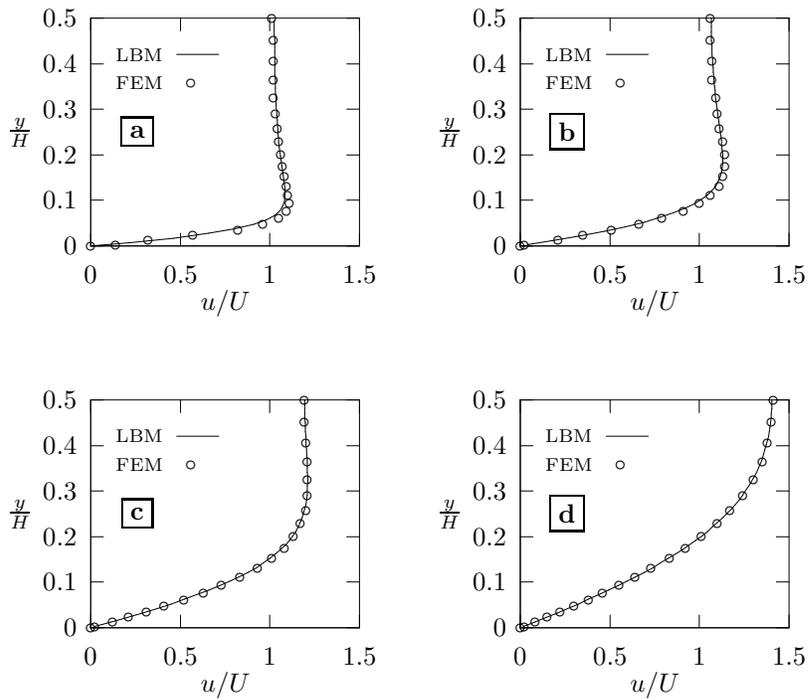

In addition, in the boundary layer theory it is assumed that for incompressible
flow the velocity profiles cannot have overshoots within the boundary layer but
the results obtained with the Finite Element Method for entrance-region flow
\cite{b11} do not confirm this assumption.
Fig. \ref{fig4} presents the velocity profile at four different distances from inlet
section, obtained with LBM and FEM, respectively.

\begin{figure}
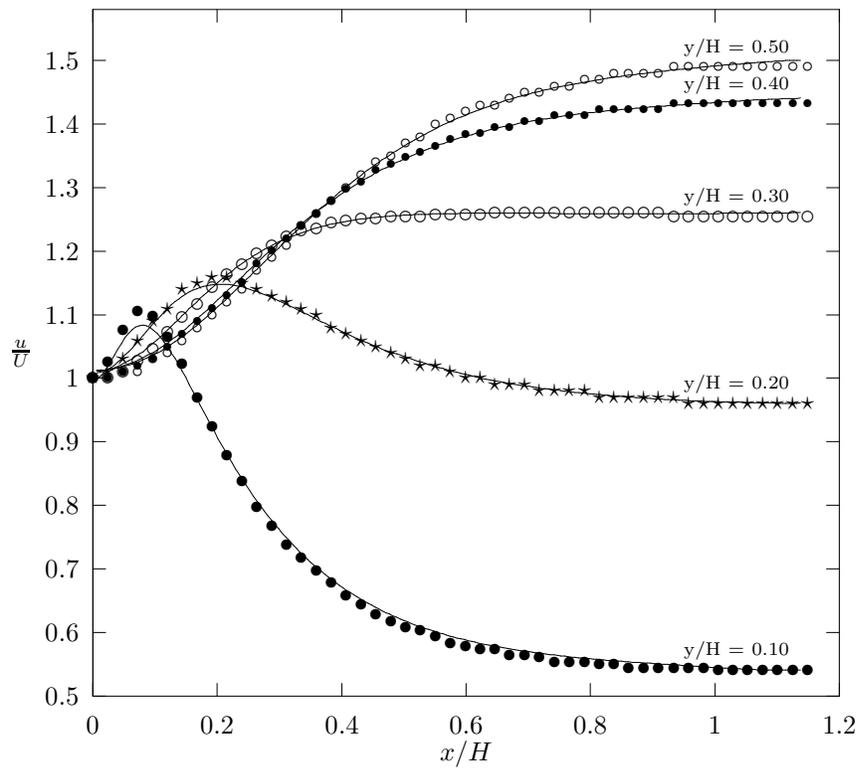

\begin{center}
\caption{\label{fig5}
Velocity  variation along the channel at
different distances from the lower plate
(solid lines -- LBM results, dots -- FEM solution). }
\setlength{\unitlength}{0.240900pt}
\ifx\plotpoint\undefined\newsavebox{\plotpoint}\fi
\sbox{\plotpoint}{\rule[-0.175pt]{0.350pt}{0.350pt}}%
\special{em:linewidth 0.3pt}%

\end{center}
\end{figure}

In order to verify if the domain length was large enough, the dimensionless
velocity, $u/U$, variation  versus  $x/H $ is presented in Fig.\ref{fig5}.
It can be seen that the curves reach asymptotically the values corresponding to
the parabolic velocity profile (\ref{eq23}).

\subsection{Unsteady unidirectional flow}

Let us consider an unidirectional flow for which the body force vanishes
at the initial moment $t=0$.
The equation of motion is:
\begin{equation}
 \frac{\partial u}{\partial t} = \nu\,
\frac{\partial^2 u}{\partial y^2} \label{eq24}
\end{equation}
with the boundary and initial conditions:
\begin{eqnarray}
 u(0,t) = U(H,t) = 0, \nonumber \\
 u(H/2,0) = U \nonumber
\end{eqnarray}
The solution is:
\begin{equation}
 u(y,t) = U\, \exp\left( -\pi^2\frac{\nu t}{H^2}\right) \,
\sin\left( \pi \frac{y}{H} \right) \label{eq25} 
\end{equation}

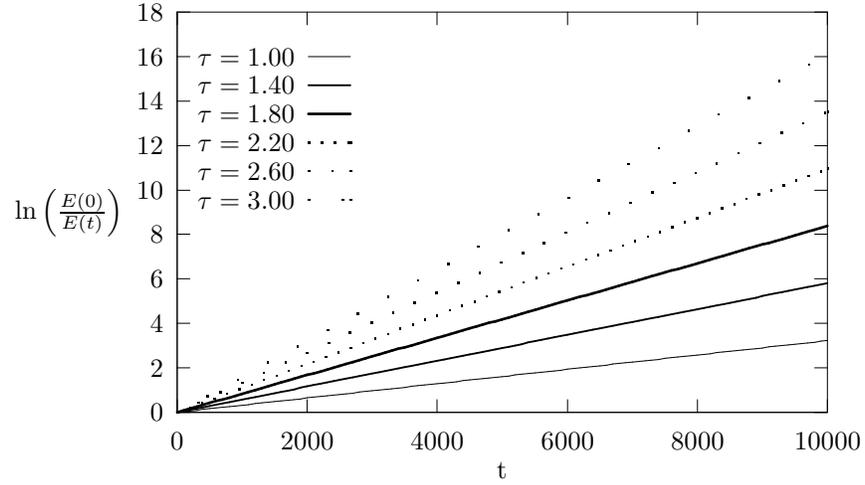
\begin{figure}
\begin{center}
\caption{\label{fig6}Kinetic energy decay.}
\setlength{\unitlength}{0.240900pt}
\ifx\plotpoint\undefined\newsavebox{\plotpoint}\fi
\sbox{\plotpoint}{\rule[-0.175pt]{0.350pt}{0.350pt}}%
\special{em:linewidth 0.3pt}%
\begin{picture}(1349,900)(0,0)
\tenrm
\put(264,158){\special{em:moveto}}
\put(1285,158){\special{em:lineto}}
\put(264,158){\special{em:moveto}}
\put(264,787){\special{em:lineto}}
\put(264,158){\special{em:moveto}}
\put(284,158){\special{em:lineto}}
\put(1285,158){\special{em:moveto}}
\put(1265,158){\special{em:lineto}}
\put(242,158){\makebox(0,0)[r]{0}}
\put(264,228){\special{em:moveto}}
\put(284,228){\special{em:lineto}}
\put(1285,228){\special{em:moveto}}
\put(1265,228){\special{em:lineto}}
\put(242,228){\makebox(0,0)[r]{2}}
\put(264,298){\special{em:moveto}}
\put(284,298){\special{em:lineto}}
\put(1285,298){\special{em:moveto}}
\put(1265,298){\special{em:lineto}}
\put(242,298){\makebox(0,0)[r]{4}}
\put(264,368){\special{em:moveto}}
\put(284,368){\special{em:lineto}}
\put(1285,368){\special{em:moveto}}
\put(1265,368){\special{em:lineto}}
\put(242,368){\makebox(0,0)[r]{6}}
\put(264,438){\special{em:moveto}}
\put(284,438){\special{em:lineto}}
\put(1285,438){\special{em:moveto}}
\put(1265,438){\special{em:lineto}}
\put(242,438){\makebox(0,0)[r]{8}}
\put(264,507){\special{em:moveto}}
\put(284,507){\special{em:lineto}}
\put(1285,507){\special{em:moveto}}
\put(1265,507){\special{em:lineto}}
\put(242,507){\makebox(0,0)[r]{10}}
\put(264,577){\special{em:moveto}}
\put(284,577){\special{em:lineto}}
\put(1285,577){\special{em:moveto}}
\put(1265,577){\special{em:lineto}}
\put(242,577){\makebox(0,0)[r]{12}}
\put(264,647){\special{em:moveto}}
\put(284,647){\special{em:lineto}}
\put(1285,647){\special{em:moveto}}
\put(1265,647){\special{em:lineto}}
\put(242,647){\makebox(0,0)[r]{14}}
\put(264,717){\special{em:moveto}}
\put(284,717){\special{em:lineto}}
\put(1285,717){\special{em:moveto}}
\put(1265,717){\special{em:lineto}}
\put(242,717){\makebox(0,0)[r]{16}}
\put(264,787){\special{em:moveto}}
\put(284,787){\special{em:lineto}}
\put(1285,787){\special{em:moveto}}
\put(1265,787){\special{em:lineto}}
\put(242,787){\makebox(0,0)[r]{18}}
\put(264,158){\special{em:moveto}}
\put(264,178){\special{em:lineto}}
\put(264,787){\special{em:moveto}}
\put(264,767){\special{em:lineto}}
\put(264,113){\makebox(0,0){0}}
\put(468,158){\special{em:moveto}}
\put(468,178){\special{em:lineto}}
\put(468,787){\special{em:moveto}}
\put(468,767){\special{em:lineto}}
\put(468,113){\makebox(0,0){2000}}
\put(672,158){\special{em:moveto}}
\put(672,178){\special{em:lineto}}
\put(672,787){\special{em:moveto}}
\put(672,767){\special{em:lineto}}
\put(672,113){\makebox(0,0){4000}}
\put(877,158){\special{em:moveto}}
\put(877,178){\special{em:lineto}}
\put(877,787){\special{em:moveto}}
\put(877,767){\special{em:lineto}}
\put(877,113){\makebox(0,0){6000}}
\put(1081,158){\special{em:moveto}}
\put(1081,178){\special{em:lineto}}
\put(1081,787){\special{em:moveto}}
\put(1081,767){\special{em:lineto}}
\put(1081,113){\makebox(0,0){8000}}
\put(1285,158){\special{em:moveto}}
\put(1285,178){\special{em:lineto}}
\put(1285,787){\special{em:moveto}}
\put(1285,767){\special{em:lineto}}
\put(1285,113){\makebox(0,0){10000}}
\put(264,158){\special{em:moveto}}
\put(1285,158){\special{em:lineto}}
\put(1285,787){\special{em:lineto}}
\put(264,787){\special{em:lineto}}
\put(264,158){\special{em:lineto}}
\put(774,68){\makebox(0,0){t}}
\put(9,472){\makebox(0,0)[l]{$\ln \left( \frac{E(0)}{E(t)} \right)$}}
\put(448,717){\makebox(0,0)[r]{$\tau = 1.00$}}
\put(470,717){\special{em:moveto}}
\put(536,717){\special{em:lineto}}
\put(264,158){\special{em:moveto}}
\put(274,159){\special{em:lineto}}
\put(284,160){\special{em:lineto}}
\put(295,161){\special{em:lineto}}
\put(305,163){\special{em:lineto}}
\put(315,164){\special{em:lineto}}
\put(325,165){\special{em:lineto}}
\put(335,166){\special{em:lineto}}
\put(346,167){\special{em:lineto}}
\put(356,168){\special{em:lineto}}
\put(366,169){\special{em:lineto}}
\put(376,170){\special{em:lineto}}
\put(387,172){\special{em:lineto}}
\put(397,173){\special{em:lineto}}
\put(407,174){\special{em:lineto}}
\put(417,175){\special{em:lineto}}
\put(427,176){\special{em:lineto}}
\put(438,177){\special{em:lineto}}
\put(448,178){\special{em:lineto}}
\put(458,179){\special{em:lineto}}
\put(468,181){\special{em:lineto}}
\put(478,182){\special{em:lineto}}
\put(489,183){\special{em:lineto}}
\put(499,184){\special{em:lineto}}
\put(509,185){\special{em:lineto}}
\put(519,186){\special{em:lineto}}
\put(529,187){\special{em:lineto}}
\put(540,188){\special{em:lineto}}
\put(550,190){\special{em:lineto}}
\put(560,191){\special{em:lineto}}
\put(570,192){\special{em:lineto}}
\put(581,193){\special{em:lineto}}
\put(591,194){\special{em:lineto}}
\put(601,195){\special{em:lineto}}
\put(611,196){\special{em:lineto}}
\put(621,197){\special{em:lineto}}
\put(632,199){\special{em:lineto}}
\put(642,200){\special{em:lineto}}
\put(652,201){\special{em:lineto}}
\put(662,202){\special{em:lineto}}
\put(672,203){\special{em:lineto}}
\put(683,204){\special{em:lineto}}
\put(693,205){\special{em:lineto}}
\put(703,206){\special{em:lineto}}
\put(713,208){\special{em:lineto}}
\put(723,209){\special{em:lineto}}
\put(734,210){\special{em:lineto}}
\put(744,211){\special{em:lineto}}
\put(754,212){\special{em:lineto}}
\put(764,213){\special{em:lineto}}
\put(774,214){\special{em:lineto}}
\put(785,215){\special{em:lineto}}
\put(795,217){\special{em:lineto}}
\put(805,218){\special{em:lineto}}
\put(815,219){\special{em:lineto}}
\put(826,220){\special{em:lineto}}
\put(836,221){\special{em:lineto}}
\put(846,222){\special{em:lineto}}
\put(856,223){\special{em:lineto}}
\put(866,224){\special{em:lineto}}
\put(877,226){\special{em:lineto}}
\put(887,227){\special{em:lineto}}
\put(897,228){\special{em:lineto}}
\put(907,229){\special{em:lineto}}
\put(917,230){\special{em:lineto}}
\put(928,231){\special{em:lineto}}
\put(938,232){\special{em:lineto}}
\put(948,234){\special{em:lineto}}
\put(958,235){\special{em:lineto}}
\put(968,236){\special{em:lineto}}
\put(979,237){\special{em:lineto}}
\put(989,238){\special{em:lineto}}
\put(999,239){\special{em:lineto}}
\put(1009,240){\special{em:lineto}}
\put(1020,241){\special{em:lineto}}
\put(1030,243){\special{em:lineto}}
\put(1040,244){\special{em:lineto}}
\put(1050,245){\special{em:lineto}}
\put(1060,246){\special{em:lineto}}
\put(1071,247){\special{em:lineto}}
\put(1081,248){\special{em:lineto}}
\put(1091,249){\special{em:lineto}}
\put(1101,250){\special{em:lineto}}
\put(1111,252){\special{em:lineto}}
\put(1122,253){\special{em:lineto}}
\put(1132,254){\special{em:lineto}}
\put(1142,255){\special{em:lineto}}
\put(1152,256){\special{em:lineto}}
\put(1162,257){\special{em:lineto}}
\put(1173,258){\special{em:lineto}}
\put(1183,259){\special{em:lineto}}
\put(1193,261){\special{em:lineto}}
\put(1203,262){\special{em:lineto}}
\put(1214,263){\special{em:lineto}}
\put(1224,264){\special{em:lineto}}
\put(1234,265){\special{em:lineto}}
\put(1244,266){\special{em:lineto}}
\put(1254,267){\special{em:lineto}}
\put(1265,268){\special{em:lineto}}
\put(1275,270){\special{em:lineto}}
\put(1285,271){\special{em:lineto}}
\sbox{\plotpoint}{\rule[-0.350pt]{0.700pt}{0.700pt}}%
\special{em:linewidth 0.7pt}%
\put(448,672){\makebox(0,0)[r]{$\tau = 1.40$}}
\put(470,672){\special{em:moveto}}
\put(536,672){\special{em:lineto}}
\put(264,158){\special{em:moveto}}
\put(274,160){\special{em:lineto}}
\put(284,162){\special{em:lineto}}
\put(295,164){\special{em:lineto}}
\put(305,166){\special{em:lineto}}
\put(315,168){\special{em:lineto}}
\put(325,170){\special{em:lineto}}
\put(335,172){\special{em:lineto}}
\put(346,174){\special{em:lineto}}
\put(356,176){\special{em:lineto}}
\put(366,178){\special{em:lineto}}
\put(376,180){\special{em:lineto}}
\put(387,182){\special{em:lineto}}
\put(397,184){\special{em:lineto}}
\put(407,186){\special{em:lineto}}
\put(417,188){\special{em:lineto}}
\put(427,190){\special{em:lineto}}
\put(438,192){\special{em:lineto}}
\put(448,194){\special{em:lineto}}
\put(458,197){\special{em:lineto}}
\put(468,199){\special{em:lineto}}
\put(478,201){\special{em:lineto}}
\put(489,203){\special{em:lineto}}
\put(499,205){\special{em:lineto}}
\put(509,207){\special{em:lineto}}
\put(519,209){\special{em:lineto}}
\put(529,211){\special{em:lineto}}
\put(540,213){\special{em:lineto}}
\put(550,215){\special{em:lineto}}
\put(560,217){\special{em:lineto}}
\put(570,219){\special{em:lineto}}
\put(581,221){\special{em:lineto}}
\put(591,223){\special{em:lineto}}
\put(601,225){\special{em:lineto}}
\put(611,227){\special{em:lineto}}
\put(621,229){\special{em:lineto}}
\put(632,231){\special{em:lineto}}
\put(642,233){\special{em:lineto}}
\put(652,235){\special{em:lineto}}
\put(662,237){\special{em:lineto}}
\put(672,239){\special{em:lineto}}
\put(683,241){\special{em:lineto}}
\put(693,243){\special{em:lineto}}
\put(703,245){\special{em:lineto}}
\put(713,247){\special{em:lineto}}
\put(723,249){\special{em:lineto}}
\put(734,251){\special{em:lineto}}
\put(744,253){\special{em:lineto}}
\put(754,255){\special{em:lineto}}
\put(764,257){\special{em:lineto}}
\put(774,259){\special{em:lineto}}
\put(785,261){\special{em:lineto}}
\put(795,263){\special{em:lineto}}
\put(805,265){\special{em:lineto}}
\put(815,268){\special{em:lineto}}
\put(826,270){\special{em:lineto}}
\put(836,272){\special{em:lineto}}
\put(846,274){\special{em:lineto}}
\put(856,276){\special{em:lineto}}
\put(866,278){\special{em:lineto}}
\put(877,280){\special{em:lineto}}
\put(887,282){\special{em:lineto}}
\put(897,284){\special{em:lineto}}
\put(907,286){\special{em:lineto}}
\put(917,288){\special{em:lineto}}
\put(928,290){\special{em:lineto}}
\put(938,292){\special{em:lineto}}
\put(948,294){\special{em:lineto}}
\put(958,296){\special{em:lineto}}
\put(968,298){\special{em:lineto}}
\put(979,300){\special{em:lineto}}
\put(989,302){\special{em:lineto}}
\put(999,304){\special{em:lineto}}
\put(1009,306){\special{em:lineto}}
\put(1020,308){\special{em:lineto}}
\put(1030,310){\special{em:lineto}}
\put(1040,312){\special{em:lineto}}
\put(1050,314){\special{em:lineto}}
\put(1060,316){\special{em:lineto}}
\put(1071,318){\special{em:lineto}}
\put(1081,320){\special{em:lineto}}
\put(1091,322){\special{em:lineto}}
\put(1101,324){\special{em:lineto}}
\put(1111,326){\special{em:lineto}}
\put(1122,328){\special{em:lineto}}
\put(1132,330){\special{em:lineto}}
\put(1142,332){\special{em:lineto}}
\put(1152,334){\special{em:lineto}}
\put(1162,336){\special{em:lineto}}
\put(1173,338){\special{em:lineto}}
\put(1183,341){\special{em:lineto}}
\put(1193,343){\special{em:lineto}}
\put(1203,345){\special{em:lineto}}
\put(1214,347){\special{em:lineto}}
\put(1224,349){\special{em:lineto}}
\put(1234,351){\special{em:lineto}}
\put(1244,353){\special{em:lineto}}
\put(1254,355){\special{em:lineto}}
\put(1265,357){\special{em:lineto}}
\put(1275,359){\special{em:lineto}}
\put(1285,361){\special{em:lineto}}
\sbox{\plotpoint}{\rule[-0.500pt]{1.000pt}{1.000pt}}%
\special{em:linewidth 1.0pt}%
\put(448,627){\makebox(0,0)[r]{$\tau = 1.80$}}
\put(470,627){\special{em:moveto}}
\put(536,627){\special{em:lineto}}
\put(264,158){\special{em:moveto}}
\put(274,161){\special{em:lineto}}
\put(284,164){\special{em:lineto}}
\put(295,167){\special{em:lineto}}
\put(305,170){\special{em:lineto}}
\put(315,173){\special{em:lineto}}
\put(325,176){\special{em:lineto}}
\put(335,178){\special{em:lineto}}
\put(346,181){\special{em:lineto}}
\put(356,184){\special{em:lineto}}
\put(366,187){\special{em:lineto}}
\put(376,190){\special{em:lineto}}
\put(387,193){\special{em:lineto}}
\put(397,196){\special{em:lineto}}
\put(407,199){\special{em:lineto}}
\put(417,202){\special{em:lineto}}
\put(427,205){\special{em:lineto}}
\put(438,208){\special{em:lineto}}
\put(448,211){\special{em:lineto}}
\put(458,214){\special{em:lineto}}
\put(468,217){\special{em:lineto}}
\put(478,219){\special{em:lineto}}
\put(489,222){\special{em:lineto}}
\put(499,225){\special{em:lineto}}
\put(509,228){\special{em:lineto}}
\put(519,231){\special{em:lineto}}
\put(529,234){\special{em:lineto}}
\put(540,237){\special{em:lineto}}
\put(550,240){\special{em:lineto}}
\put(560,243){\special{em:lineto}}
\put(570,246){\special{em:lineto}}
\put(581,249){\special{em:lineto}}
\put(591,252){\special{em:lineto}}
\put(601,255){\special{em:lineto}}
\put(611,258){\special{em:lineto}}
\put(621,260){\special{em:lineto}}
\put(632,263){\special{em:lineto}}
\put(642,266){\special{em:lineto}}
\put(652,269){\special{em:lineto}}
\put(662,272){\special{em:lineto}}
\put(672,275){\special{em:lineto}}
\put(683,278){\special{em:lineto}}
\put(693,281){\special{em:lineto}}
\put(703,284){\special{em:lineto}}
\put(713,287){\special{em:lineto}}
\put(723,290){\special{em:lineto}}
\put(734,293){\special{em:lineto}}
\put(744,296){\special{em:lineto}}
\put(754,299){\special{em:lineto}}
\put(764,301){\special{em:lineto}}
\put(774,304){\special{em:lineto}}
\put(785,307){\special{em:lineto}}
\put(795,310){\special{em:lineto}}
\put(805,313){\special{em:lineto}}
\put(815,316){\special{em:lineto}}
\put(826,319){\special{em:lineto}}
\put(836,322){\special{em:lineto}}
\put(846,325){\special{em:lineto}}
\put(856,328){\special{em:lineto}}
\put(866,331){\special{em:lineto}}
\put(877,334){\special{em:lineto}}
\put(887,337){\special{em:lineto}}
\put(897,340){\special{em:lineto}}
\put(907,342){\special{em:lineto}}
\put(917,345){\special{em:lineto}}
\put(928,348){\special{em:lineto}}
\put(938,351){\special{em:lineto}}
\put(948,354){\special{em:lineto}}
\put(958,357){\special{em:lineto}}
\put(968,360){\special{em:lineto}}
\put(979,363){\special{em:lineto}}
\put(989,366){\special{em:lineto}}
\put(999,369){\special{em:lineto}}
\put(1009,372){\special{em:lineto}}
\put(1020,375){\special{em:lineto}}
\put(1030,378){\special{em:lineto}}
\put(1040,381){\special{em:lineto}}
\put(1050,383){\special{em:lineto}}
\put(1060,386){\special{em:lineto}}
\put(1071,389){\special{em:lineto}}
\put(1081,392){\special{em:lineto}}
\put(1091,395){\special{em:lineto}}
\put(1101,398){\special{em:lineto}}
\put(1111,401){\special{em:lineto}}
\put(1122,404){\special{em:lineto}}
\put(1132,407){\special{em:lineto}}
\put(1142,410){\special{em:lineto}}
\put(1152,413){\special{em:lineto}}
\put(1162,416){\special{em:lineto}}
\put(1173,419){\special{em:lineto}}
\put(1183,422){\special{em:lineto}}
\put(1193,424){\special{em:lineto}}
\put(1203,427){\special{em:lineto}}
\put(1214,430){\special{em:lineto}}
\put(1224,433){\special{em:lineto}}
\put(1234,436){\special{em:lineto}}
\put(1244,439){\special{em:lineto}}
\put(1254,442){\special{em:lineto}}
\put(1265,445){\special{em:lineto}}
\put(1275,448){\special{em:lineto}}
\put(1285,451){\special{em:lineto}}
\sbox{\plotpoint}{\rule[-0.250pt]{0.500pt}{0.500pt}}%
\special{em:linewidth 0.5pt}%
\put(448,582){\makebox(0,0)[r]{$\tau = 2.20$}}
\put(470,582){\usebox{\plotpoint}}
\put(490,582){\usebox{\plotpoint}}
\put(511,582){\usebox{\plotpoint}}
\put(532,582){\usebox{\plotpoint}}
\put(536,582){\usebox{\plotpoint}}
\put(264,158){\usebox{\plotpoint}}
\put(264,158){\usebox{\plotpoint}}
\put(283,165){\usebox{\plotpoint}}
\put(302,172){\usebox{\plotpoint}}
\put(322,179){\usebox{\plotpoint}}
\put(341,187){\usebox{\plotpoint}}
\put(361,194){\usebox{\plotpoint}}
\put(380,201){\usebox{\plotpoint}}
\put(399,209){\usebox{\plotpoint}}
\put(419,215){\usebox{\plotpoint}}
\put(438,223){\usebox{\plotpoint}}
\put(458,231){\usebox{\plotpoint}}
\put(477,237){\usebox{\plotpoint}}
\put(497,245){\usebox{\plotpoint}}
\put(516,252){\usebox{\plotpoint}}
\put(536,259){\usebox{\plotpoint}}
\put(555,267){\usebox{\plotpoint}}
\put(574,274){\usebox{\plotpoint}}
\put(594,281){\usebox{\plotpoint}}
\put(613,289){\usebox{\plotpoint}}
\put(633,296){\usebox{\plotpoint}}
\put(652,303){\usebox{\plotpoint}}
\put(671,310){\usebox{\plotpoint}}
\put(691,318){\usebox{\plotpoint}}
\put(710,325){\usebox{\plotpoint}}
\put(730,332){\usebox{\plotpoint}}
\put(749,340){\usebox{\plotpoint}}
\put(768,347){\usebox{\plotpoint}}
\put(788,354){\usebox{\plotpoint}}
\put(807,362){\usebox{\plotpoint}}
\put(827,369){\usebox{\plotpoint}}
\put(846,376){\usebox{\plotpoint}}
\put(866,384){\usebox{\plotpoint}}
\put(885,390){\usebox{\plotpoint}}
\put(905,398){\usebox{\plotpoint}}
\put(924,405){\usebox{\plotpoint}}
\put(943,412){\usebox{\plotpoint}}
\put(963,420){\usebox{\plotpoint}}
\put(982,427){\usebox{\plotpoint}}
\put(1002,434){\usebox{\plotpoint}}
\put(1021,441){\usebox{\plotpoint}}
\put(1040,449){\usebox{\plotpoint}}
\put(1060,457){\usebox{\plotpoint}}
\put(1079,463){\usebox{\plotpoint}}
\put(1099,471){\usebox{\plotpoint}}
\put(1118,478){\usebox{\plotpoint}}
\put(1138,485){\usebox{\plotpoint}}
\put(1157,493){\usebox{\plotpoint}}
\put(1176,500){\usebox{\plotpoint}}
\put(1196,507){\usebox{\plotpoint}}
\put(1215,514){\usebox{\plotpoint}}
\put(1235,522){\usebox{\plotpoint}}
\put(1254,529){\usebox{\plotpoint}}
\put(1274,536){\usebox{\plotpoint}}
\put(1285,541){\usebox{\plotpoint}}
\put(448,537){\makebox(0,0)[r]{$\tau = 2.60$}}
\put(470,537){\usebox{\plotpoint}}
\put(507,537){\usebox{\plotpoint}}
\put(536,537){\usebox{\plotpoint}}
\put(264,158){\usebox{\plotpoint}}
\put(264,158){\usebox{\plotpoint}}
\put(297,173){\usebox{\plotpoint}}
\put(331,189){\usebox{\plotpoint}}
\put(365,204){\usebox{\plotpoint}}
\put(399,220){\usebox{\plotpoint}}
\put(433,236){\usebox{\plotpoint}}
\put(467,251){\usebox{\plotpoint}}
\put(501,267){\usebox{\plotpoint}}
\put(534,283){\usebox{\plotpoint}}
\put(568,299){\usebox{\plotpoint}}
\put(602,314){\usebox{\plotpoint}}
\put(636,330){\usebox{\plotpoint}}
\put(670,346){\usebox{\plotpoint}}
\put(704,361){\usebox{\plotpoint}}
\put(738,377){\usebox{\plotpoint}}
\put(772,393){\usebox{\plotpoint}}
\put(806,408){\usebox{\plotpoint}}
\put(839,424){\usebox{\plotpoint}}
\put(873,440){\usebox{\plotpoint}}
\put(907,456){\usebox{\plotpoint}}
\put(941,471){\usebox{\plotpoint}}
\put(975,487){\usebox{\plotpoint}}
\put(1009,503){\usebox{\plotpoint}}
\put(1043,518){\usebox{\plotpoint}}
\put(1077,534){\usebox{\plotpoint}}
\put(1110,549){\usebox{\plotpoint}}
\put(1144,565){\usebox{\plotpoint}}
\put(1178,581){\usebox{\plotpoint}}
\put(1212,597){\usebox{\plotpoint}}
\put(1246,612){\usebox{\plotpoint}}
\put(1280,628){\usebox{\plotpoint}}
\put(1285,630){\usebox{\plotpoint}}
\put(448,492){\makebox(0,0)[r]{$\tau = 3.00$}}
\put(470,492){\usebox{\plotpoint}}
\put(523,492){\usebox{\plotpoint}}
\put(536,492){\usebox{\plotpoint}}
\put(264,158){\usebox{\plotpoint}}
\put(264,158){\usebox{\plotpoint}}
\put(311,183){\usebox{\plotpoint}}
\put(358,209){\usebox{\plotpoint}}
\put(405,236){\usebox{\plotpoint}}
\put(452,261){\usebox{\plotpoint}}
\put(500,287){\usebox{\plotpoint}}
\put(547,313){\usebox{\plotpoint}}
\put(594,339){\usebox{\plotpoint}}
\put(641,365){\usebox{\plotpoint}}
\put(689,391){\usebox{\plotpoint}}
\put(736,418){\usebox{\plotpoint}}
\put(783,444){\usebox{\plotpoint}}
\put(830,469){\usebox{\plotpoint}}
\put(878,495){\usebox{\plotpoint}}
\put(925,522){\usebox{\plotpoint}}
\put(972,548){\usebox{\plotpoint}}
\put(1019,573){\usebox{\plotpoint}}
\put(1066,600){\usebox{\plotpoint}}
\put(1113,626){\usebox{\plotpoint}}
\put(1161,652){\usebox{\plotpoint}}
\put(1208,678){\usebox{\plotpoint}}
\put(1255,704){\usebox{\plotpoint}}
\end{picture}
\end{center}
\end{figure}

The free decay of the total fluid kinematic energy can be expressed as:
\begin{equation}
 \frac{E_c(0)}{E_c(t)} = \exp\left( 2\pi^2 \frac{\nu t}{H^2}\right) \label{eq26}
\end{equation}
and it can be used to calculate the kinematic viscosity from the slope of
the $ln\left[ E_c(0)/E_c(t)\right]\leftrightarrow t $ line.
Such lines are presented in Fig.\ref{fig6}, the numerical results being obtained for
a lattice with $N_w = 101 $ and $N_l = 100 $.

Table \ref{tab2} presents the numerical and analytical values for $\nu $, the relative
difference being below 0.2\%.

\begin{table}
\caption{\label{tab2}
Kinematic viscosities ($\nu_{an}$ from 
(\protect\ref{eq14}) and $\nu_{num}$ obtained
from the kinetic energy decay) for $0.6 \leq \tau \leq 3.0 $. }
\begin{center}
\begin{tabular}{cccc}
$\tau$ & $\nu_{an}$ & $\nu_{num}$ & $ \varepsilon_{\nu}$ [\%] \\
\hline
$ 0.60 $  &  $  2.50 \cdot 10^{-2} $ &  $  2.50023 \cdot 10^{-2} $ & $  0.0092 $ \\
$ 0.80 $  &  $  7.50 \cdot 10^{-2} $ &  $  7.50041 \cdot 10^{-2} $ & $  0.0055 $ \\ 
$ 1.00 $  &  $  1.25 \cdot 10^{-1} $ &  $  1.25000 \cdot 10^{-1} $ & $  0.0000 $ \\
$ 1.20 $  &  $  1.75 \cdot 10^{-1} $ &  $  1.74986 \cdot 10^{-1} $ & $  0.0079 $ \\
$ 1.40 $  &  $  2.25 \cdot 10^{-1} $ &  $  2.24958 \cdot 10^{-1} $ & $  0.0184 $ \\
$ 1.60 $  &  $  2.75 \cdot 10^{-1} $ &  $  2.74906 \cdot 10^{-1} $ & $  0.0340 $ \\
$ 1.80 $  &  $  3.25 \cdot 10^{-1} $ &  $  3.24846 \cdot 10^{-1} $ & $  0.0471 $ \\
$ 2.00 $  &  $  3.75 \cdot 10^{-1} $ &  $  3.74754 \cdot 10^{-1} $ & $  0.0654 $ \\
$ 2.20 $  &  $  4.25 \cdot 10^{-1} $ &  $  4.24633 \cdot 10^{-1} $ & $  0.0862 $ \\
$ 2.40 $  &  $  4.75 \cdot 10^{-1} $ &  $  4.74479 \cdot 10^{-1} $ & $  0.1096 $ \\
$ 2.60 $  &  $  5.25 \cdot 10^{-1} $ &  $  5.24287 \cdot 10^{-1} $ & $  0.1357 $ \\
$ 2.80 $  &  $  5.75 \cdot 10^{-1} $ &  $  5.74054 \cdot 10^{-1} $ & $  0.1644 $ \\
$ 3.00 $  &  $  6.25 \cdot 10^{-1} $ &  $  6.23777 \cdot 10^{-1} $ & $  0.1957 $  \\
\end{tabular}
\end{center}
\end{table}

\section{Time development of steady flow between parallel plates
         in relative motion}

Suppose now that the fluid and the two parallel plates are
initially at rest state. After the lower plate is suddenly brought 
to the steady velocity U in its own plane, while
the upper one is still maintained at rest, 
the governing differential equation is (\ref{eq24}) again,
with the boundary and initial conditions
\begin{eqnarray}
 u(0,t) = U, \quad u(H,t) & = & 0, \quad {\rm for} \quad t>0  \nonumber \\
 u(y,0) & = & 0, \quad {\rm for }\quad 0<y\le H \nonumber
\end{eqnarray}
The velocity distribution is given by \cite{b8} :
\begin{equation}
 u(y,t) = U\left( 1 - \frac{y}{H} \right)
\frac{2U}{\pi} \sum_{i=1}^{\infty} \frac{1}{i}\,\exp\left( -i^2\pi^2\frac{\nu t}{H^2}\right)
\, \sin\left(i\pi \frac{y}{H}\right) \label{eq27} 
\end{equation}

The LBM simulation was made for a lattice with $N_w = 101 $ and $N_l = 100 $.
The distance between plates was now 
$H\, = \,101\,\sqrt{3}/2\, - \, \sqrt{3}/4 $. The term $\sqrt{3}/4$ 
comes from the fact that we imposed the velocity $U=0.01 $ on the lower
plate (there was no "bounce-back" condition there).

\begin{figure}
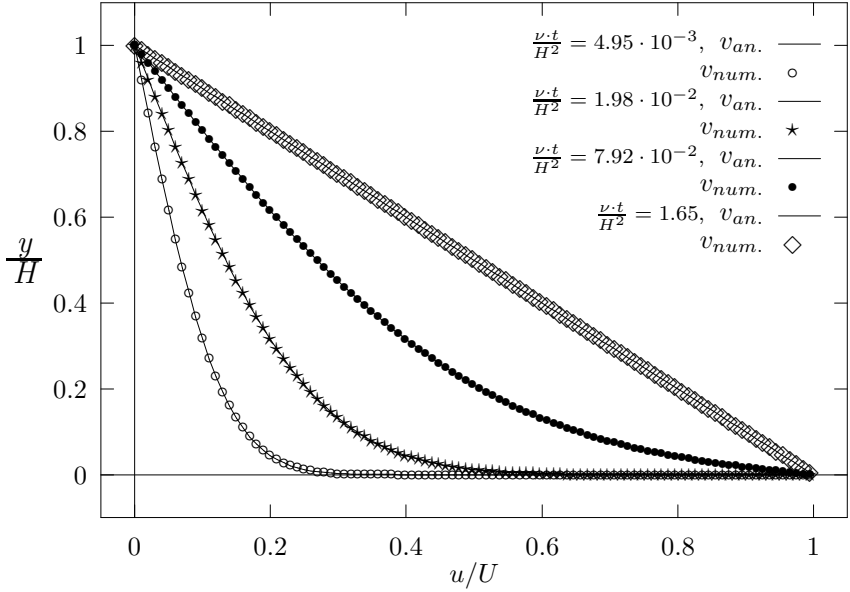

\begin{center}
\caption{\label{fig7}
Development of velocity profile between
parallel plates in relative motion
without pressure gradient
(dots -- LBM results, solid lines --
analytical solution, eq.\protect\ref{eq27}). }
\setlength{\unitlength}{0.240900pt}
\ifx\plotpoint\undefined\newsavebox{\plotpoint}\fi
\sbox{\plotpoint}{\rule[-0.175pt]{0.350pt}{0.350pt}}%
\special{em:linewidth 0.3pt}%

\end{center}
\end{figure}

The time evolution of the velocity profile is represented 
in Fig. \ref{fig7}, at $t =$ 300, 1200, 4800 and 10000,
the kinematic viscosity being $\nu = 0.125 $.

An excellent  concordance can be observed between the numerical
and analytical results.
The linear velocity profile corresponding to the
steady Couette flow with no pressure gradient 
it is naturally obtained for $t\rightarrow\infty $.

\section{Conclusions}

A numerical approach to four kinds of flow between two parallel plates 
using the Lattice Boltzmann Method is presented. 
The number of lattice nodes was chosen in order achieve an acceptable CPU
time on Alfa DEC computers.
The numerical and analytical values of the kinematic viscosity were used to
estimate the accuracy of the LBM simulations for both steady and unsteady
unidirectional flows, the relative differences being found below one percent.

The zero velocity conditions on 
the rest channel walls were imposed by using "bounce-back" rules.
The moving plate condition was achieved by  forcing the 
the equilibrium distribution
functions for the corresponding nodes in order to  
satisfy the imposed velocity.

Since the Schlichting approximate analytical solution for the entrance-region
flow is inadequate at low Reynolds numbers,
the comparison of the LBM results with a Finite Element Method 
solution was found to be appropriate.

Finally, we found an excellent 
agreement between the LBM computed time evolution
of the velocity profiles and the analytical solutions 
of the Couette flow without pressure gradient.

\section*{Acknowledgments}
The financial support of the Ministry of Education (contract 3012 / 1994)
and of the Ministry of Research and Technology (contract 475 B / 1994)
is gratefully acknowledged.

\end{document}